\documentclass[12pt]{article}
\usepackage{mydef}
\usepackage{authblk}
\usepackage{epsfig}

\usepackage[hide]{ed}

\begin{document}

\title{USLV: Unspanned Stochastic Local Volatility Model
\thanks{Opinions expressed in this paper are those of the authors, and do not necessarily reflect the view of JPMorgan Chase and Numerix.}}
\author[1,2]{Igor Halperin}
\author[3,2]{Andrey Itkin}

\affil[1]{{\small MR\&D, JPMorgan Chase, 270 Park Avenue, New York, NY 10172, USA\\igor.halperin@jpmorgan.com}}
\affil[2]{\small Polytechnic Institute of New York University, 6 Metro Tech Center, RH 517E, Brooklyn NY 11201, USA}
\affil[3]{\small Numerix LLC, 150 East 42nd Street, 15th Floor, New York, NY 10017, USA\\aitkin@numerix.edu}

\date{\today}

\maketitle

\begin{abstract}
We propose a new framework for modeling stochastic local volatility, with potential
applications to modeling derivatives on interest rates, commodities, credit, equity, FX etc.,
as well as hybrid derivatives. Our model extends
the linearity-generating unspanned volatility term structure model by Carr et al.\ (2011)
by adding a local volatility layer to it. We outline efficient numerical schemes for
pricing derivatives in this framework for a particular four-factor specification
(two ``curve'' factors plus two ``volatility'' factors). We show that the dynamics of such a
system can be approximated by a Markov chain on a two-dimensional space $(Z_t,Y_t)$, where coordinates $ Z_t $ and $Y_t $ are given by direct (Kroneker) products of values of
pairs of curve and volatility factors, respectively. The resulting Markov chain dynamics
on such partly ``folded'' state space enables fast pricing by the standard backward induction.
Using a nonparametric specification of the Markov chain generator, one can
accurately match arbitrary sets of vanilla option quotes with different strikes and maturities.
Furthermore, we consider an alternative formulation of the model in terms of an implied time
change process. The latter is specified nonparametrically, again enabling accurate calibration
to arbitrary sets of vanilla option quotes.

\end{abstract}

\vspace{0.5in}

\section{Introduction}

\subsection{Motivation}

The present work is motivated by the desire to
have a unified modeling methodology and shared implementation for derivatives
pricing with a dynamic volatility smile for various asset classes, including
interest rates (IR), commodities, equities,
credit, foreign exchange (FX) etc., as well as for modeling hybrid derivatives such as
equity-IR or
equity-commodities hybrids. We present one possible approach, which extends a recently proposed class
of stochastic volatility models.

\subsection{Related Previous Work}

\citet{Gabaix_2007} proposed a new class of asset price models, the so-called
{\it linearity-generating processes} (LGP). Such processes are defined by the condition
that the current prices of basic instruments (stock, bonds, futures, swaps etc.)
are {\it linear} in a driving Markov process $ X_t $. This stands in sharp contrast
to popular {\it affine} models where, e.g.,  a zero-coupon bond price $ P(t,T) $
is an exponentially-affine function of a Markov driver $ X_t $.

On the theoretical side, the LGP processes appear very attractive. Indeed,
typical ways we model basic instruments are drastically different between, say,
IR and
equity models\footnote{Both are taken to be examples of term-structure models vs.\
spot-models. Instead of IR and equity, we could, e.g. compare commodities and FX.}. In the
equity world, basic instruments (equities) are linear in stochastic factors (usually taken to be
equity prices themselves for purposes of modeling derivatives), and volatility is
stochastic (SV) and unspanned (USV, see below for the definition of this term).

In the IR world, the mathematics are almost the same for the HJM-type
models that model the entire yield curve. As the yield curve is in
one-to-one correspondence with bond prices, it can be viewed as an ``observable'' basic instrument
that again is linear in factors, and typically gives rise to USV.

But such linearity of HJM-like models has a high price, namely that the number of state variables
needed for the Markovian dynamics turns out to be too high for use in a lattice-based setting
in most cases of practical interest. Therefore, even with Markovian specifications, HJM-like
models are typically
employed within a Monte Carlo setup rather than on a lattice. On the other hand, an attempt to
reduce the curve modeling to a short-rate modeling, as is done in affine models, leads to a
nonlinear relation between bond prices and the factors, which produces undesirable side effects,
such as a dependence of the instantaneous forward curve on the short-rate volatility.\footnote{This
is the cost one has to pay for non-linearity. Clearly, nothing similar ever occurs in spot
models: today's stock price $ S_t $ is obviously independent of the current volatility or current
value of the volatility factor $ Y_t $. Mathematically, this can be formulated as
the statement that for spot stochastic volatility models (such as e.g. the Heston model),
the pricing function $ f(S_t, Y_t) $ of basic instruments (stocks) is an identity
$ f(S_t, Y_t) = S_t $,
see also Table \ref{Tab} in Sect.\ \ref{sec:overview}.}

This problem is resolved in the LGP approach. By putting both equity and
bonds on equal footing in terms of making them both linear functionals of the factors
(and doing it in a different way from HJM), the LGP-based models play a role of
``grand unification models,'' similar in a conceptual sense to ``grand unification theories'' (GUT) in physics. No proliferation of the number of Markov drivers occurs in
LGP-type models as we move from one class of basic instruments (stocks) to other class (bonds).

Also on the practical side, linearity has profound consequences for tractability of
asset pricing modeling within the LGP framework. In particular, if a zero-coupon
bond price is linear in $ X_t $, then so will be prices of a coupon bond or a swap. As a result, the swaption pricing, e.g. can be done in a semi-analytical form without additional
approximations, such as those used by the Libor Market Models (LMM). It is also very helpful
in calibration, as will be discussed in more detail below.

To summarize, the class of LGP-like models identified by Gabaix is a new interesting class
that may develop into a viable competitor to both affine models, which are currently one of the main
workhorses for derivatives modeling in credit, commodities, rates and other asset classes, and also
HJM-type models.  Yet this approach is in its infancy compared to the well-studied class of affine models.

In 2011, Carr, Gabaix and Wu (CGW) proposed a LGP-type stochastic volatility term structure
model (\citet{CGW}). CGW, in particular, emphasize the point that stochastic volatility
generated in LGP-type models is {\it unspanned} in the sense of the definition of \citet{CDG},
who coined the original term ``unspanned volatility''\footnote{Following \citet{CDG}, the
volatility is called unspanned if bond prices do not depend on the stochastic volatility
driver.}. The CGW model offers a number of attractive features.  Most importantly, it is a low-dimension Markov model with unspanned stochastic volatility (USV), and an orthogonal set of
model parameters with a separate calibration to the term structure and option volatilities.

The CGW model is a pure stochastic volatility model, as volatility is modeled as a superposition of CIR processes.
To make it more practical, it would be very useful to add
a local volatility layer to the model. Our extension of the CGW model amounts to
introduction of such a local volatility factor, along with efficient numerical methods for
calibration and pricing. To differentiate our framework from CGW, in what follows
we will refer to it as the unspanned stochastic local volatility (USLV) model.

\section{Overview of Our Framework} \label{sec:overview}

By construction, USLV preserves the linearity and USV properties of the CGW approach.
Another property inherited from the CGW model is that USLV is formulated directly
in the physical measure $ \mathbb{P} $ (see below) rather than in the risk-neutral measure $ \mathbb{Q} $, which makes it easier, e.g., to combine the historical and pricing data for model estimation, if desired.

The main theoretical construction that USLV adds to the CGW model is a local volatility layer.
The resulting mixed stochastic/local volatility dynamics has a few important implications.

First, adding a local volatility layer enables nearly perfect matching of an arbitrary number of
European vanilla option quotes with different strikes and maturities.\footnote{Note
that while this property of USLV is shared by local stochastic volatility models as well, the key
point here is that now we have an additional risk factor (volatility) to acknowledge, model and
hedge.}
Such an extension is clearly
desirable in order to apply this approach for pricing of both
vanilla and exotic derivatives, especially if vanilla options are used to hedge the exotics.

Second, the presence of a local volatility layer alongside a stochastic volatility part
induces a decomposition of the option volatility into spanned and unspanned
parts, rather than being of a pure unspanned type as in the CGW model.
One could expect that such decomposition of volatility should translate into a
decomposition of an option's vega into a delta-vega and a ``genuine vega'' part.

Because of the way our model is calibrated, it enables traders to incorporate their view on the relative
weights of the spanned and unspanned parts in the option's vegas.\footnote{Technically, this is
done by giving the end user the ability to input his/her own set of speed factors (SF), see
below.} By viewing a trader's inputs as a prior model that does not necessarily match observed
options, our model finds a minimal adjustment (``tweak'') to the trader's prior model in
order to reinforce an accurate match of the option quotes.

In contrast, the volatility in local volatility models would be 100\% spanned. In local volatility models, matching
vanilla pricing would fix the volatility surface for all strikes and
maturities, and would not leave any flexibility for the model to match prices of more exotic
options. The inclusion of stochastic volatility allows one to simultaneously have more realistic
forward smile dynamics and additional parameters to match exotics'
prices (if available).  The ability of USLV to incorporate a possible trader's view is
what sets it apart from both pure local volatility models and pure SV models of the
CGW type.

On the implementation side, USLV concentrates on the most important low-dimensional
specifications for practicality, e.g., two factors for the term structure (with $ N $ curve factors in general),
and one or two factors  for stochastic volatility (with $ M $ volatility factors in general).
In particular, for a (2+2)-factor case, we show how to approximate the dynamics of the
driving factors by a two-dimensional Markov chain on a space
constructed by folding (see below) of the original four-dimensional state space. This
enables fast pricing by standard backward induction on the chain.

It should be noted that while in this paper we concentrate on modeling term-structure dynamics
(e.g., of futures, swap rates or credit spreads) with potential
applications to ``term structure asset classes,'' such as IR, commodities or credit, the same
approach can be used for modeling spot prices, which would be a proper
setting for ``spot asset classes,'' such as equities or FX. Moreover, due to a symmetric treatment of
``term-structure assets'' and ``spot assets'' in the present framework, this approach is readily available for
modeling hybrid derivative products (e.g., equity-IR or equity-commodity hybrids) using
the same implementation. Changes from one asset class to another
would amount to a proper reparametrization and reinterpretation of the Markov generator matrices while
leaving the computational algorithm intact.\footnote{In principle, this could produce a
generic pricing engine, similar in a sense to Monte Carlo (MC). Indeed, the latter method is a
``universal'' method of derivatives pricing in the sense that in this framework, we only need to implement
dynamic equations and payoff functions for a particular model-product combination in
order to use a generic MC engine. Likewise, our Markov chain framework is
``universal'' in the same sense (within a class of all diffusive
local stochastic volatility models in up to (2+2)
dimensions). The only difference here is that while in MC we typically start with continuous space
dynamics, which is then discretized for simulation through discretization of processes
(e.g., Brownian motions) driving the dynamics, the dynamics in our approach
are fundamentally defined in terms of discretized state variables.}
Furthermore, in the continuous limit, different parametrizations of
the stochastic volatility generator in our Markov chain model
give rise to a rich class of (2+2)-factor models, including stochastic local volatility with
jumps. Note that in this paper we primarily concentrate on specifications whose continuous limit
is a two-dimensional
diffusion with a two-dimensional diffusive stochastic local volatility. This case is covered in detail
below in Sect.~\ref{sect:USLV_N2_M2}. However, in Sect.~\ref{sect:USLV22_with_ITC},
we will present an alternative formulation that can give rise to jumps in both the underlying
and stochastic volatility. Our approach
is thus quite flexible in its ability to accommodate different specifications of the dynamics, including
a four-factor stochastic local volatility model with jumps.


\subsection{USLV vs.\ HJM vs.\ Affine Models}
Our initial interest in using LGP-type models for modeling stochastic volatility was
inspired by the observation that LGP-type models (and, by extension, USLV-like models) seem
to combine the best features of both HJM-type models and affine models, while avoiding
their disadvantages. Indeed, like the HJM-type models, the stochastic volatility is
unspanned in USLV. Unlike the HJM-types, the model is Markov in dimension $ N + M$ rather
than $ N + N(N+1)/2 + M $, as in HJM-type models. Conversely, both affine models and USLV have
the same number of state variables ($ N + M $). However, in USLV, volatility is always (partly)
unspanned, while in affine models, volatility in general will be spanned unless
some special constraints are imposed on parameters, which might be restrictive for calibration
purposes. (See also Table \ref{Tab} below.)

The above reasoning suggests that if we manage to generalize the pure stochastic volatility
model of CGW to a stochastic local volatility model (i.e., to make a USLV out of CGW), and
do it in a numerically efficient way, and if the resulting model demonstrates
good parameters and hedges stability etc., then such a model can be considered a viable
candidate for use in practice. This paper outlines the theoretical
framework for USLV, leaving numerical experiments for future work.

A few more words of caution are in order here. Our outline of the USLV is generic and
is not tied yet to any specific asset class. Each asset class makes its own demands on a model. For example, the ability
to reproduce the Samuelson effect and asset cointegration are very important for
commodities, alongside the ability to handle seasonality in asset levels and volatilities for
certain commodities, such as gas or power.  It has yet to be seen how (or whether) the USLV
framework can accommodate such specific requirements. A discussion of this matter is planned for the second stage
of the present theoretical work.

A brief summary of different model classes is presented in Table \ref{Tab},
where we compare the behavior of equity stochastic volatility models such as the Heston model,
HJM-type, affine-type and LGP/CGW/USLV-type. The third column shows the functional form of
conditional expectations arising in calculation of prices of elementary instruments.
\begin{table}
\begin{center}
\begin{tabular}{|c|c|c|c|c|c|}
\hline
Model & $ BI_t $ & $ BV_t $  & $ BI_t = f(BV_t) $ & USV  & D   \cr
\hline
 Equity & $ S_t $ & $ S_t, Y_t $ & $ f(S_t, Y_t) = S_t $ & Yes & $ 1 + M $   \cr
 \hline
 IR HJM & $ P_t^T $ & $ P_t^T, Y_t $ & $ f(P_t^T, Y_t) = P_t^T $ & Yes
 & $ N + N(N+1)/2 + M $  \cr
 \hline
 IR Affine & $ P_t^T $ & $ X_t, Y_t $ &  $ f(X_t,Y_t) $ & No  & $ N + M $
  \cr
 \hline
 CGW/USLV & $ P_t^T $ & $ X_t , Y_t $ &  $ f(X_t,Y_t) = \alpha_{tT} X_t +
 \beta_{tT} $ & Yes & $ N + M $
  \cr
 \hline
\end{tabular}
\end{center}
\caption{Model comparison summary. Note that ``Yes'' in column USV for IR HJM means
``in general, yes,'' and likewise ``No'' for IR Affine means ``in general, no, unless
special 'knife-edge' constraints are imposed on parameters of the model.'' Here $ S_t, P_t^T $
and $ Y_t $ stand for the stock and bond
prices and volatility factor, respectively, while $ BI_t $ and $ BV_t $ stand for
basic instruments and basic variables, respectively. Finally, $ D $ stands for for the total number of state
variables needed
for a Markovian description.}
\label{Tab}
\end{table}

\section{The Carr-Gabaix-Wu Model}

In this section, we provide a brief overview of the CGW model of \citet{CGW}. The
CWG model is then used as
the first step in our setting. Simultaneously, in this section we set our
notation, on which we largely follow \citet{CGW}.

\subsection{State-Price Processes and Martingale Pricing}
The famous fundamental theorem of asset pricing (\cite{HarrisonPliska}) states that if the economy is
arbitrage free, then there exists (under certain technical conditions such as positivity and time consistence) a strictly
positive process $ M_t $ called the {\it state space deflator}, such that the deflated gain
process associated with any admissible trading strategy is a martingale under the
measure $ \mathbb{P} $. In particular, for a contingent payoff $ \Pi_T $ at time $ T > t$,
its value at time $ t $ is given by the following $ \mathbb{P}$-conditional expectation:
\beq
\label{V_exp}
V(t,T) = \mathbb{E}_t \left[ \frac{M_T}{M_t} \Pi_T \right]
\eeq
The ratio $ M_T/M_t $ is sometimes referred to as the stochastic discount factor or the
pricing kernel. The $ \mathbb{P}$-measure SDE for $ M_t $ reads
\[
\frac{d M_t}{M_t} = - r dt - \boldsymbol \lambda(t, {\bf Z}_t) d {\bf Z}_t,
\]
where $ {\bf Z}_t $ is a vector of risk factors and $ \boldsymbol \lambda(t, {\bf Z}_t) $ measures
the market prices of risk for these factors.
The formal solution to this SDE takes a multiplicative form
\[
M_t = M_0 \exp \left( -  \int_{0}^{t} r_s ds \right)
\mathcal{E} \left( - \int_{0}^{t} \lambda(t,Z_s) dZ_s
\right),
\]
where $ \mathcal{E}( \cdot)  $ stands for the stochastic exponential martingale operator.
The latter defines the Radon-Nikod{\'y}m derivative $ d \mathbb{Q} / d \mathbb{P} $ that
transforms the physical measure $ \mathbb{P} $ to the risk-neutral measure $
\mathbb{Q} $ such that, under $ \mathbb{Q}$, the contingent claim valuation reads
\beq
\label{Q_measure}
V(t,T) = \mathbb{E}_t^{\mathbb{Q}} \left[ \exp \left( - \int_{t}^{T} r_s ds \right) \Pi_T \right].
\eeq

\subsection{One-Factor Case}

Our starting point is a version of one-factor LGP dynamics suggested
in \cite{BB}. It differs from \cite{CGW} by i) parametrization, and ii) derivation and interpretation of the main result of the LGP approach (see \eqref{PDE_bond_sol} below).
The two formulations can, however, be mapped onto one another by a proper re-parametrization.

Assume that a state variable $ X_t $ is driven by the
following SDE under measure $ \mathbb{P}$:
\beq
\label{X_state}
dX_t = r_t X_t dt + \sigma(t,X_t) \left[ \lambda(t,X_t)dt + dW_t \right],
\eeq
where $ W_t  $ is the standard Brownian motion and the
short rate $ r_t $ is an affine function of $ X_t $:
\beq
\label{r_t_N=1}
r_t = a - c e^{- \mu t} X_t \equiv a - c \tilde{X}_t
\eeq
where $ \tilde{X}_t = e^{ - \mu t} X_t $ is a detrended state variable with $ \mu $ being
the growth rate of $ X_t $, and
$ a > 0 $ and $ c > 0 $ are two constants\footnote{Note that we can
set $ c = 1 $ without any loss of generality. However, we prefer to keep it in
the formulae for correct dimensionality, i.e. $ c $ has
the dimensionality of $ r_t $ divided by that of $ X_t $.}.

The $ \mathbb{P}$-measure SDE for the state price deflator $ M_t $ reads
\beq
\label{SDE_SPD}
\frac{d M_t}{M_t} = - r_t dt - \lambda(t,X_t) dW_t
\eeq
The representation of stochastic dynamics given in \eqref{X_state} and \eqref{SDE_SPD}
is convenient as both the real world (measure $ \mathbb{P}$) and
risk-neutral (measure $ \mathbb{Q} $) dynamics can be directly read from
these equations provided we know functions $ \sigma(t,X_t)$ and $ \lambda(t,X_t) $.
Indeed, by the Girsanov theorem, the combination $ dW_t + \lambda(t,X_t) dt $ is
a $ \mathbb{Q} $-martingale, so that the $ \mathbb{Q}$-measure drift of
$ X_t $ is the short rate $ r_t $. This means that the state variable
$ X_t$ can
be interpreted economically as a risky (tradable or non-tradable) asset, or
be
related to aggregate market returns, see \cite{BB}.

Using \eqref{SDE_SPD} and the fact that $ M_t X_t $ is a martingale under measure
$ \mathbb{P} $ (which is easy to verify using It{\'o}'s lemma and \eqref{X_state} and
\eqref{SDE_SPD}), we
compute the conditional time-$t $ expectation of $ M_T $ as follows\footnote{All conditional
expectations $ \mathbb{E}_t \left[ \cdot \right] $  in this paper refer to measure $ \mathbb{P} $
unless specifically stated otherwise.}:
\bea
 \mathbb{E}_t \left[ M_T \right] &=&
\mathbb{E}_t \left[ M_t - \int_{t}^{T} r_s M_s ds -
\int_{t}^{T} \lambda(s,X_s) M_s d W_s \right] \nonumber \\
 &=& M_t - \int_{t}^{T} a \mathbb{E}_t \left[ M_s \right] ds
+ \int_{t}^{T} c e^{-\mu s}  \mathbb{E}_t \left[ M_s X_s \right] ds
 \\
 &=& M_t - \int_{t}^{T} a \mathbb{E}_t \left[ M_s \right] ds
+ \left( \int_{t}^{T} c e^{-\mu s} ds \right) M_t X_t \nonumber
\eea
This produces the following result for the time-$t $ zero-coupon bond price
\beq
\label{BP_1factor}
P(t,T) = \frac{\mathbb{E}_t \left[ M_T \right]}{M_t}
= 1 - \int_{t}^{T} a P(t,s) ds +
\left( \int_{t}^{T} c e^{-\mu s} ds \right) X_t
\eeq
Differentiating this relation with respect to $ T $, we transform it
into a differential equation:
\beq
\label{PDE_bond}
\frac{\partial P(t,T)}{\partial T} = - a P(t,T) + c e^{-\mu T} X_t
\eeq
whose solution reads (\cite{BB})
\beq
\label{PDE_bond_sol}
P(t,T) = e^{-a \tau} \left[ 1 + c e^{-\mu t} \frac{e^{(a-\mu) \tau} - 1}{
a - \mu} X_t \right] = e^{-a \tau} \left[ 1 + c
\frac{e^{(a-\mu) \tau} - 1}{
a - \mu} \tilde{X}_t \right]\, , \; \; \tau = T-t
\eeq
Note that this expression does not depend on specification
of functions $ \sigma(t,X_t) $ and $ \lambda(t,X_t) $ in \eqref{X_state},
and the shape of the yield curve at time $ t $ is driven
by the current value of $ X_t $. This is different from
affine models where the bond price is represented as a conditional expectation of
a non-linear functional of future values of the state variables.

\begin{figure}[ht]
\begin{center}
\fbox{\includegraphics[width=4.5 in]{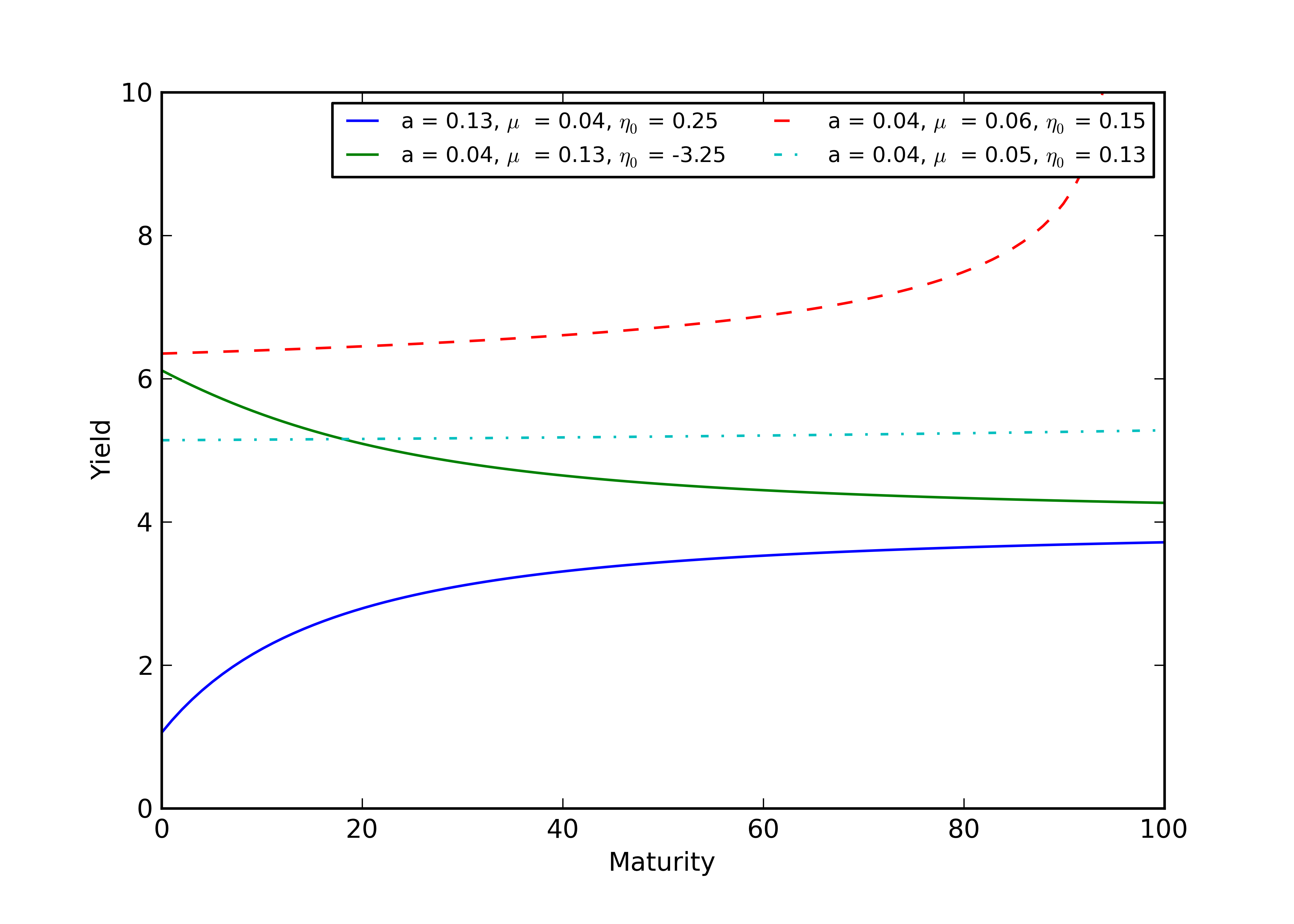}}
\caption{Yield curves obtained with a one-factor specification
\eqref{PDE_bond_sol} with $ t = 0 $ for
different combinations of parameters $ a, \, \mu $ and $ \eta_0 $, as a function of
maturity in years.
Parameter $ c $ equals 0.1 for all curves.}
\label{fig:yield_crvs_1_factor}
\end{center}
\end{figure}

The solution of \eqref{X_state} in the zero volatility limit
$ \sigma(t,X_t) = 0 $ reads
\bea \label{zero_vol_sol}
X_t = \left\{ \begin{array}{lll}
 \dfrac{a-\mu}{c} \frac{e^{\mu t} }{1 - \eta_0 e^{-(a-\mu)t} },
& \eta_0 = 1 - \frac{a-\mu}{c X_0}, & \mbox{if } \mu \neq a,  \\
\dfrac{e^{\mu t}}{\eta_0 + c t}, & \eta_0 = \frac{1}{X_0}, & \mbox{if } \mu = a
\end{array}
\right.
\eea
In what follows, we concentrate on the case $ \mu < a $ which
corresponds to
the scenario where the detrended asset $ \tilde{X}_t = e^{-\mu t} X_t $
reaches
a constant level $ \tilde{X}_{\infty} =   \frac{a-\mu}{c} > 0 $ as
$ t \rightarrow \infty $.
In addition, we assume that
$ X_0 >  \frac{a- \mu}{c} $, so that $ 0< \eta_0 < 1 $ which produces an
increasing and concave yield curve, while negative values
of $ \eta_0 $ produce inverted yield curves, see
Fig.~\ref{fig:yield_crvs_1_factor}.
In this scenario,
the short rate  $ r_t $ approaches $ \mu $ as  $ t \rightarrow \infty $.
If $ \eta_0 $ satisfies a more restrictive constraint
$ \eta_0 < \mu/a$, then
the short rate $ r_t $ stays positive (or equivalently $ P(t,T) \leq 1 $)
for all $ t $.
Therefore, we assume a slightly more general constraint on $ \eta_0 $ by
introducing
another parameter $ \bar{\mu} $ that lies in the interval $ [\mu, a] $:
\beq
\label{eta_0}
0 \leq \eta_0 < \frac{\bar{\mu}}{a} \; , \; \; \mu \leq \bar{\mu} \leq a
\eeq
For such choice of $ \eta_0 $, \eqref{zero_vol_sol} with $ \mu < a $ produces a
concave and monotonically increasing yield curve, while the
short rate $ r_t $ and state process $ X_t $
are bounded at all times as follows:
\beq
\label{bounds_rX}
a - \frac{a - \mu}{1 - \left( \bar{\mu}/a \right) e^{-(a-\mu)t}}  \leq r_t \leq \mu, \qquad
\frac{a - \mu}{c} \leq \tilde{X}_t \leq  \frac{ a - \mu}{c} \frac{1}{1 - \left(\bar{\mu}/a \right) e^{-(a-\mu)t}}
\eeq
where the low and upper boundaries for $ r_t $ (respectively, upper and lower boundaries
for $ \tilde{X}_t $)  are attained by setting $ \eta_0 = \bar{\mu} /a $ or $ \eta_0 = 0 $,
respectively. One sees that parameter $ \bar{\mu} $ can be used to control how much the
short rate $ r_t $ can go negative. If we set $ \bar{\mu} = \mu $, the short rate $ r_t $ is non-negative
at all times $ t \geq 0 $, but as we will see shortly such
choice for $ \bar{\mu} $ may not necessarily be optimal in the present
framework.

Now we return to \eqref{X_state} with non-vanishing volatility, and
consider the following ansatz for $ X_t $:
\beq
\label{X_t_ansatz}
X_t = \frac{a-\mu}{c} \frac{e^{\mu t}}{1 - Y_t e^{-(a-\mu)t} } \, , \; \;  \mu < a
\eeq
which is obtained by replacing a constant $ \eta_0 $ in \eqref{zero_vol_sol} by
a stochastic process $ Y_t $ restricted to the strip $ [0, \bar{\mu} / a] $.
Such a process with a given initial value $ 0 \leq Y_0 \leq \bar{\mu}/a $ can be constructed as follows:
\beq
\label{Y_from_Z}
Y_t =  \frac{ \bar{\mu} Y_0}{a Y_0  + ( \bar{\mu} - a Y_0) Z_t}
\eeq
where $ Z_t \geq 0 $ is a
non-negative martingale under measure $ \mathbb{P} $
that starts at
$ Z_0 = 1 $ and satisfies the following SDE:
\beq
\label{Z_t_SDE}
\frac{d Z_t}{Z_t} =  - \hat{\sigma}(t,Z_t) dW_t
\eeq
where $ \hat{\sigma}(t,Z_t) $ is a local volatility function that
will be specified later.
The initial value $ X_0 $ is expressed in terms of $ Y_0 $ as follows:
\beq
\label{X_0}
X_0 = \frac{a - \mu}{c} \frac{1}{1 - Y_0}
\eeq
Substituting \eqref{Y_from_Z} into \eqref{X_t_ansatz}, we obtain
\beq
\label{X_from_Z}
X_{t} = \frac{a - \mu}{c}
\frac{ e^{ \mu t} \left[ a Y_0 + \left( \bar{\mu} - a Y_0 \right) Z_t \right] }{
a Y_0 \left[ 1 - (\bar{\mu}/a) e^{-(a - \mu)t} \right] + \left(
\bar{\mu} - a Y_0 \right) Z_t }
\equiv
\frac{a - \mu}{c}e^{ \mu t}
\frac{ \left( \alpha + \beta Z_t \right)}{ \alpha \kappa(t) + \beta_0 Z_t}
\eeq
where
\beq
\label{alpha_kappa}
\alpha = a Y_0 \; , \; \; \beta = \bar{\mu} - \alpha \; , \; \; \kappa(t) = 1 -
\frac{ \bar{\mu}}{a} e^{-(a - \mu)t}
\eeq
\eqref{X_from_Z} defines $ X_t $ as a fractional-linear
function\footnote{This function is
also known as a M{\"o}bius transformation in
complex analysis.} of the martingale process $ Z_t $ determined
by \eqref{Z_t_SDE}. Clearly, the process $ X_t $ defined by \eqref{X_from_Z} is
bounded, as it should be as long as it is related by a linear equation
\eqref{PDE_bond_sol} to the bond price which {\it is} bounded\footnote{One
can notice here a certain conceptual similarity between the LGP and Markov
functional
models on the one hand, and a difference with the affine models on the other hand.
For the latter, the driving SDE has affine drift and diffusion
coefficients, while the function $ f(X_t, Y_t) $ (see Table \ref{Tab}) is {\it nonlinear} (nonaffine).
For the LGP-type models, e.g. the CGW model, the situation is reversed: the
state equation is now
{\it nonaffine} but the pricing equation is
{\it affine}.}.
By taking
the limits $ Z_t \rightarrow 0 $ and $ Z_t \rightarrow \infty $,
one obtains
bounds for $ \tilde{X}_t $:
\beq
\label{bounds_tildeX}
\frac{a - \mu}{c} \leq \tilde{X}_t \leq \frac{ a - \mu}{c} \frac{1}{1 -
(\bar{\mu}/a) e^{-(a-\mu)t}}
\eeq
which coincides with \eqref{bounds_rX}.
Note that \cite{CGW} obtain a fractional-linear expression
for $ X_t $ as a function of $ Z_t $ that is similar to our
\eqref{X_from_Z},
however their choice of time-dependent coefficients is different.

Using Ito's lemma, we find that $ X_t $ defined
by \eqref{X_from_Z} satisfies \eqref{X_state}
provided we choose the following specification for functions
$ \sigma(t,X_t) $ and $ \lambda(t,X_t) $ (which we express here
as functions of $ Z_t $ rather than $ X_t $):
\bea
\label{sigma_lambda}
\sigma(t,X_t) &=& \alpha   \hat{\sigma}(t,Z_t)
\frac{\bar{\mu}}{a}  \frac{a- \mu}{c}
   \frac{\beta Z_t e^{-(a-2 \mu) t}}{ \left[ \alpha \kappa(t) + \beta Z_t \right]^2}
  \nonumber \\
\lambda(t,X_t) &=&  \hat{\sigma}(t,Z_t)  \frac{\beta Z_t}{ \alpha \kappa(t) + \beta Z_t }
\eea
Note that volatility $ \sigma(t, X_t) $ vanishes in both limits
$ Z_t  \rightarrow 0 $ or $ Z_t \rightarrow \infty $
which correspond to
boundaries in the $ X $-space given by \eqref{bounds_tildeX}.
Finally, the solution of \eqref{SDE_SPD} reads
\beq
\label{M_t_sol}
M_t = \frac{ e^{ - \mu t}
\left[ \alpha \kappa(t) + \beta Z_t \right]}{\alpha + \beta}
\eeq
where we normalize $M_t $ to enforce the constraint $ M_0 = 1 $.
Therefore, the state price deflator $ M_t $ is a linear functional of the
Markov driver $ Z_t $. Linearity of this expression
is exactly the reason we do not have any convexity
corrections in
\eqref{PDE_bond_sol}. For any nonlinear functional a resulting
expression for $ P(t,T) $
would depend
on volatility of $ Z_t $. Such behavior is intentionally avoided in
the present framework. Further note that  using
\eqref{M_t_sol}, \eqref{X_from_Z} can be re-written as follows:
\beq
\label{X_from_Z_2}
X_{t} = \frac{a - \mu}{c ( \alpha + \beta)} \frac{ \alpha + \beta Z_t}{M_t}
\eeq
Therefore,
for contingent payoffs $ \Pi_T $ that are linear in $ X_T $, the
conditional expectation in \eqref{V_exp} reduces to
the expectation of a linear functional of
$ Z_T $\footnote{In addition, \eqref{X_from_Z_2} shows self-consistency of our approach: while
we know from \eqref{X_state} and \eqref{SDE_SPD}
that
the product $ M_t X_t $ is a martingale, \eqref{X_from_Z_2} shows that
this product is proportional to an affine function
$ \alpha + \beta Z_t $ of a martingale $ Z_t $, which is a
martingale.}.
This property of the
model can be
used, in particular, in order to reduce the swaption pricing to a single option on a
future value $ Z_T $, see \cite{CGW} for more details.

\begin{figure}[ht]
\begin{center}
\fbox{\includegraphics[width=4.5 in]{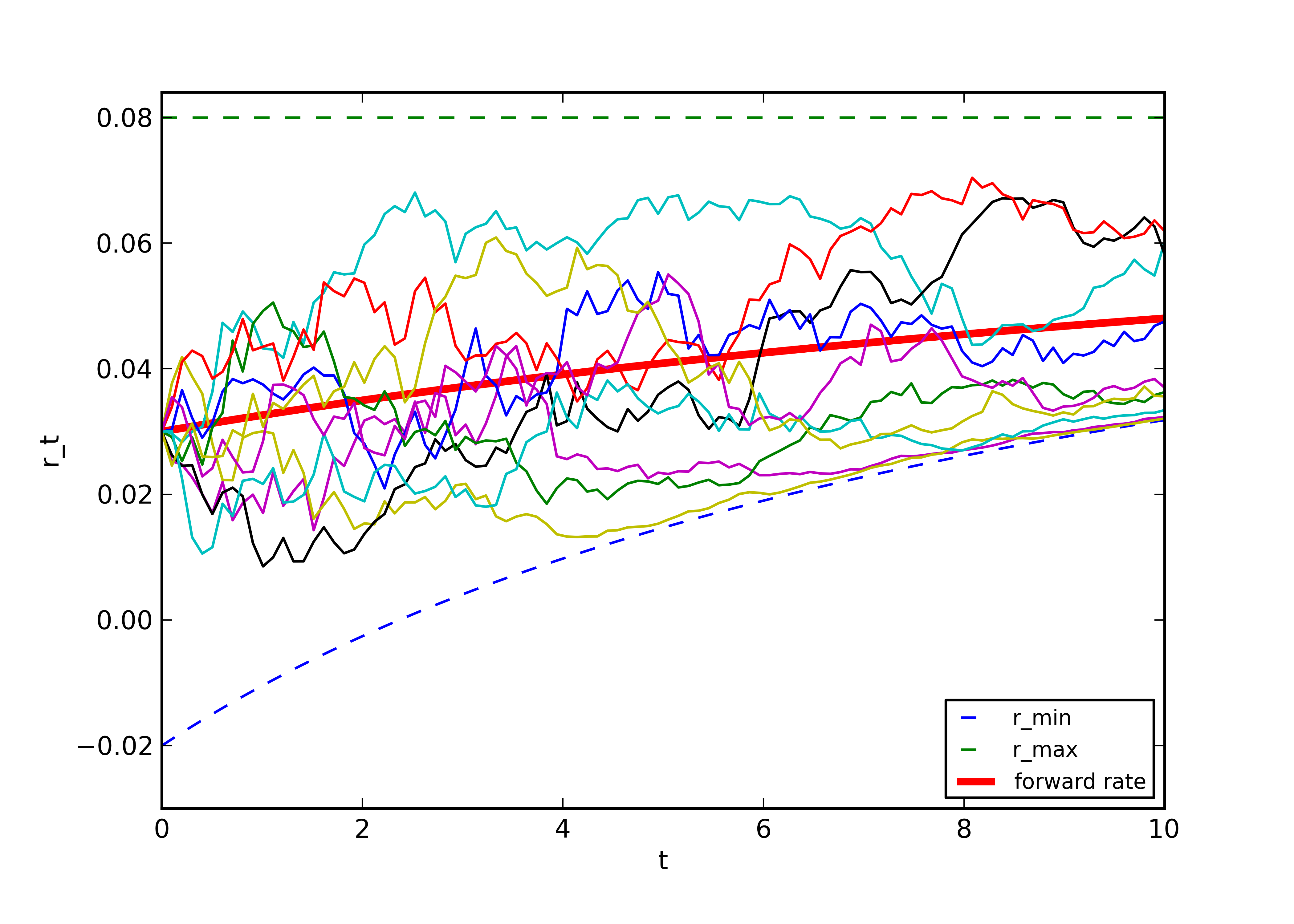}}
\caption{Simulated paths of the short rate $ r_t $ as a function of calendar time $ t $ (in years)
for the one-factor LGP process \eqref{X_from_Z},
for the following
choice of parameters: $ a = 0.085 $,  
$ \mu = 0.080 $, $ \bar{\mu} = 0.081 $, $ Y_0 = 0.909 $
and $ \hat{\sigma} = 0.6 $.}
\label{fig:Simulated_r_t}
\end{center}
\end{figure}

Note that as $ t \rightarrow \infty $, the
detrended process $ \tilde{X}_t = e^{ - \mu t} X_t $ with $ X_t $
defined by \eqref{X_from_Z} converges to the following equilibrium value:
\beq
\label{tilde_X_asym}
\tilde{X}_{\infty} = \frac{a - \mu}{c}
\eeq
irrespective of the behavior of the process $ Z_t $. In other words, stochasticity
dies off in the one-factor LGP model, while
the onset of this
asymptotic regime
is attained at times $ t $ where $ \kappa(t) $ approaches
unity:
\beq
\label{t_asym}
\frac{ \bar{\mu}}{a} e^{ - (a - \mu) t} \ll 1  \; \Leftrightarrow \; t \gg \frac{1}{a-\mu}
\eeq
The resulting asymptotic extinction of volatility of the process
$ X_t $ as $t \rightarrow \infty $
appears to be a theoretical limitation of a one-factor LGP framework.
This behavior is illustrated in Fig.~\ref{fig:Simulated_r_t}
where we show a few paths of a simulated short rate process.
As will be shown in the next section, a two-factor formulation
allows one to build dynamics
where stochasticity does not die off as $ t \rightarrow \infty $.
Another limitation of the one-factor formulation is that it is only able
to produce monotonic (increasing or decreasing) yield curves.
In order to get a humped-shaped yield curve, one needs at least
a two factor specification.

On the practical side, the effect
of decaying stochasticity of $ X_t $ in the one-factor formulation can
be somewhat mitigated by a suitable choice of parameters.
In particular, if the yield curve is not too steep, by taking $ \bar{\mu} > \mu $,
we can admit some negative short rates in the short term while guaranteeing that
$ r_t $ stays positive for longer maturities, as shown in
Fig.~\ref{fig:Simulated_r_t} where $ r_t $ can go negative for $ t \leq 2 $ years.
Such simple trick allows one to somewhat delay shrinkage of the support region
of $ r_t $ for larger time horizons.
Therefore, we expect
that the one-factor version of the model can still be used for
monotonic yield curves with sufficiently short time horizons.



\subsection{Extension to a Multi-Factor Case}
The previous formulation treats a one-factor case $ N = 1 $.
For an arbitrary number of factors $ N \geq 1 $, the state vector
$ {\bf X}_t = \left(X_{1t}, \ldots, X_{Nt} \right) $ satisfies
the following equation:
\beq
\label{X_state_N=2}
d {\bf X}_t = r_t {\bf X}_t dt + \boldsymbol \sigma(t,{\bf X}_t)
\left[ \boldsymbol \lambda(t, {\bf X}_t)dt + d {\bf W}_t \right],
\eeq
where $ {\bf W}_t  $ is a $N$-dimensional standard Brownian motion and the
short rate $ r_t $ is an affine function of $ {\bf X}_t $:
\beq
\label{r_t_N=2}
r_t = a - c \sum_{i=1}^{N}  e^{- \mu_i t} X_{it} \equiv
a - c \sum_{i=1}^{N}  \tilde{X}_{it}
\eeq
where $ \tilde{X}_{it} = e^{ - \mu_i t} X_{it} $ are detrended state
variables with $ \mu_i $ being the corresponding growth rates, and
$ a > 0 $ and $ c  > 0 $ are non-negative constants.
The $ \mathbb{P}$-measure SDE for the state price deflator $ M_t $ reads
\beq
\label{SDE_SPD_N=2}
\frac{d M_t}{M_t} = - r_t dt - \boldsymbol \lambda(t,X_t) d {\bf W}_t
\eeq
The solution to this equation generalizes \eqref{M_t_sol}:
\beq
\label{M_t_sol_N=2}
M_t = \frac{ \sum_{i=1}^{N} e^{ - \mu_i t}
\left[ \alpha_i \kappa_i(t) + \beta_i Z_t \right]}{\sum_{i=1}^{N} (\alpha_i + \beta_i)}
\eeq
where parameters $ \alpha_i, \beta_i $ and functions $ \kappa_i(t) $ with
$ i = 1, \ldots, N $ are defined similarly to \eqref{alpha_kappa}. Finally,
the solution for the bond price in a $ N $-factor model generalizes \eqref{PDE_bond_sol} (\cite{BB}):
\beq
\label{PDE_bond_sol_N=2}
P(t,T) = e^{-a \tau} \left[ 1 + c \sum_{i=1}^{N}
e^{-\mu_i t} \frac{e^{(a-\mu_i) \tau} - 1}{
a - \mu_i} X_{it} \right] = e^{-a \tau} \left[ 1 + c
\sum_{i=1}^{N} \frac{e^{(a-\mu_i) \tau} - 1}{
a - \mu_i} \tilde{X}_{it} \right]
\eeq
where $ \tau = T - t  $. As can be seen from
\eqref{PDE_bond_sol_N=2}, only the factor with the smallest value of $ \mu $
contributes to this expression in the
limit $ T \rightarrow \infty $.
In our analysis below we set $ N = 2 $ and assume
that $ \mu_2 < \mu_1 $. Therefore, we interpret
the second factor $ X_{2t} $
as a long term factor, while $ X_{1t} $ serves as a short-term factor
that drives, jointly with $ X_{2t} $, the short-term part of the
yield curve. A two-factor specification gives rise to a richer set of
yield curves than a one-factor one, including, in particular, humped curves or
curves with sign-changing convexity which could not be
obtained with a one-factor specification.

Another key difference of a multi-factor case from a one-factor one is
that volatility does not necessarily die off in the long
run. We can illustrate
this for $ N = 2 $.
Proceeding similarly to the one-factor case, we first solve the system
\eqref{X_state_N=2} without noise, by setting $ \boldsymbol \sigma(
t,{\bf X}_t) = 0 $. We obtain
\bea
\label{X_t_sol_N=2}
\tilde{X}_{1t} &=&  \frac{a - \mu_1}{c}
\frac{e^{-(\mu_1 - a)t}}{\xi_0  -  \eta_0 e^{-(\mu_2 -a)t}  + e^{-(\mu_1 - a)t}} \nonumber \\
\tilde{X}_{2t} &=&  \eta_0 \frac{\mu_2- a}{c}
\frac{e^{-(\mu_2 - a)t}}{\xi_0   - \eta_0 e^{-(\mu_2 -a)t}  + e^{-(\mu_1 - a)t}}
\eea
where $ \eta_0 $ and $ \xi_0 $ are two constants that are determined by
initial conditions. In what follows, we assume
that $ \mu_1 > a $ and $ a < \mu_2 < \mu_1 $,
as suggested by a numerical example below.
For such a scenario, the
following constraints on parameters $ \eta_0 $ and $ \xi_0 $:
\beq
0 \leq \eta_0 \leq 1 \, , \; \; \xi_0 \geq 1 
\eeq
guarantee that \eqref{X_t_sol_N=2} is well defined at all times.
Additional constraints on parameters $ \eta_0 $ and $ \xi_0 $ arise if we
enforce positivity of the short rate $ r_t = a - c \tilde{X}_{1t} - c \tilde{X}_{2t} $.
We will omit details that can be reproduced using similar steps to the
above one-factor case,
and instead concentrate on showing that volatility in a two-factor
(or, more generally, a multi-factor) LGP-type model does not in general
die off in the long term.

\begin{figure}[ht]
\begin{center}
\fbox{\includegraphics[width=4 in]{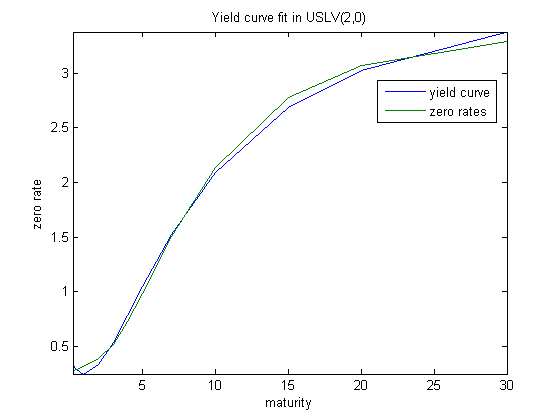}}
\caption{Calibrated yield curve (as a function of maturity in years) on 03/15/13 in the two factor model.
Calibrated parameters are as follows:
$ a=0.041, \, \mu_1 = 0.520, \, \mu_2 = 0.382, \,  X_1 = - 0.151, \,  X_2 = 0.188 $.}
\label{fig:yield_crv_2_factor}
\end{center}
\end{figure}

To this end, we note that a solution of a full $ N = 2 $ model
with non-vanishing volatility
$ \boldsymbol \sigma(t, {\bf X}_t) \neq 0 $ can be obtained along similar
lines to the previous $ N = 1 $ case.
More specifically, we should replace constants $ \xi_0 $ and
$ \eta_0 $ in the deterministic solution \eqref{X_t_sol_N=2}
by judiciously chosen fractional-linear functions of a two-dimensional
martingale process $ {\bf Z}_t = \left( Z_{1t}, Z_{2t} \right) $.
The latter satisfies a two-factor generalization of \eqref{Z_t_SDE}, and
has the initial value $ {\bf Z}_0 = (1,1) $.
On the other hand, assuming that $ \mu_1 > a $ and $ a < \mu_2 < \mu_1 $,
the asymptotic behavior of the deterministic solution
\eqref{X_t_sol_N=2} as $ t \rightarrow \infty $
is as follows:
\beq
\label{asympt_N=2_1}
\tilde{X}_{1t} \rightarrow  \frac{a - \mu_1}{c} \frac{1}{\xi_0} e^{-(\mu_1 -
a)t} \; , \; \;
 \tilde{X}_{2t} \rightarrow \frac{\mu_2-a}{c} \frac{\eta_0}{\xi_0}
e^{-(\mu_2 - a)t}
\eeq
This suggests that volatility of factors
$ X_{1t} \, X_{2t} $ does not necessarily get extinct in the long run,
unlike the behavior observed in the one-factor case. An
example of a calibrated yield curve is shown in
Fig.~\ref{fig:yield_crv_2_factor}.

\section{The USLV Model} \label{sect:USLV}
Until this point, our formalism has largely followed the CGW model by \citet{CGW} (albeit using the formulation of LGP dynamics proposed by \cite{BB}). Now it is time to part ways.

CGW investigate a parametric stochastic
volatility (SV) specification of the dynamics of $ Z_t $.  In that specification, the
stock volatility is obtained by a
stochastic time change with an activity rate process given by a superposition
of CIR processes.

Our plan is different. We want to stick to
simple one- or two-factor specifications of the stochastic volatility process, while concentrating on
modeling a nonparametric local volatility layer in \eqref{Z_t_SDE}
in such a way that all observable option quotes would be
exactly matched. Furthermore, our approach is necessarily numerical and is based on
a Markov chain approximation to the dynamics of the martingale $ Z_t $.

We will construct our model in two steps. In the first step, we develop a discretized
nonparametric local
volatility version of USLV that corresponds to a zero vol-of-vol limit of the full-blown
model.
Calibration to option quotes in this framework is achieved via a set of multiplicative adjustment
factors acting on elements of the Markov generator (see below). We will refer to these
adjustment factors as the one-dimensional (1D) {\it Speed Factors} (SFs).  Calibration of such a
one-dimensional USLV model amounts to computing a set of 1D SFs.

In the second step, we move on to a full-blown USLV model with a nonzero vol-of-vol
by turning stochastic volatility on.
In the present discrete-space framework, this amounts to
making the Markov chain generator stochastic.

To retain a near-perfect calibration to a set of
option quotes obtained in the first (1D) step, we introduce another set of speed factors
which we will refer to as 2D speed factors (2D SFs). The 2D SFs are then calibrated from
the previously computed 1D  SFs using a version of the Markovian projection method
implemented in an
efficient manner using forward induction on the Markov chain.
Once calibrated, the resulting 2D Markov chain can be utilized to set up efficient pricing
schemes for
derivatives based on backward-induction algorithms.

We note that the resulting discrete-space continuous-time dynamics on a Markov chain with
a stochastic generator arising in our approach
resembles the BSLP model developed for portfolio credit derivatives in \citet{AH}. The two
models
are similar in that both use a two-step approach to calibration. In addition,  both the
definition and  parametrization of our speed factors are similar to how analogously defined
contagion factors are introduced and used in the BSLP model.
The {\it difference} of the USLV model from the BSLP model is that while a discrete-space
description
is {\it exact} for credit,\footnote{Provided some additional assumptions are made, such
as a discrete spectrum of
recovery values.} it is used as an {\it approximation} to the dynamics of the underlying  for
USLV.
Furthermore, while
the BSLP model uses a non-linear {\it death} process as a model for a portfolio loss, in the present
case we use a nonlinear {\it quasi-birth-death} (QBD) process as a discrete approximation to the dynamics
of a two-dimensional Markov driver $ {\bf Z_t} = \left(Z_t^{(1)}, Z_t^{(2)} \right)^\top $. Finally,
we use a different method for an efficient computation of matrix
exponentials arising in evaluation probabilities on Markov chains.

In what follows, we use the following compact notation for different flavors of our model.
We denote one- and two-factor discretized local volatility models as USLV(1,0) and USLV(2,0),
respectively. Versions with stochastic volatility are denoted as USLV(1,1), (2,1), or (2,2).
We call a discretized process for $ {\bf Z_t} = \left(Z_t^{(1)}, Z_t^{(2)}\right)^\top $ a 1D process, while
the joint process of $ {\bf Z_t} $ and stochastic volatility is referred to as a 2D process.

\subsection{USLV(1,0): One-Factor Local Volatility} \label{sect:N1_USLV}
We consider the following SDE describing a local volatility dynamics of a one-dimensional Markov driver $ Z_t $ under the $ \mathbb{P} $-measure:
\beq
\label{dZ_t}
\frac{dZ_t}{Z_t} = \hat{\sigma}(Z_t, t) dW_t,
\eeq
where $ W_t $ is a Brownian motion and $ \hat{\sigma}(Z_t, t)$ is a local volatility.
Our objective is to discretize the dynamics corresponding to \eqref{dZ_t}.

To this end, we first construct a nonuniform grid of possible values of the martingale $ Z_t > 0 $ with
$ Z_0 = 1 $. Let $ p $ be the number of points on the grid and $0 < z_0 < z_1 < \cdots < z_{p-1}$ be the
nodes on the grid. Irregularity of the grid allows making it denser in interesting regions, and sparser in
uninteresting ones. Clearly, the fact that our process is a martingale helps as our grid
should not be too large: as time passes, the underlying stays around the current value in
the sense of expectations.

Let the current state be $ z_i $ and let $ \Delta z_{i,i-1} = z_i - z_{i-1}, \ i \in [1,p-1]$ be the $i$th space interval on the grid. We construct the elements of the generator matrix $ A_t $ following the
adaptive Markov chain approximation of \citet{CLS}. Adapting their
formulae to our case of a zero-drift diffusion $ Z_t $, we obtain
the Markov chain generator
\bea \label{empir_matrix}
A \, = \,  \left(
 \begin{array}{ccccc}
a_{00} & a_{01} & 0                 &  \cdots          & 0  \\
a_{10} & a_{11} & a_{12}         & \cdots           & 0 \\
\vdots  \\
0         & \cdots  & a_{p-2,p-3}  & a_{p-2,p-2}  & a_{p-2,p-1}           \\  	
0         & \cdots  & 0                  & a_{p-1,p-2}  & a_{p-1,p-1}
\end{array} \right)
\eea
with elements
\begin{align} \label{A_1_elements}
a_{i,i-1} &= \frac{ s_i(t)}{ \Delta z_{i,i-1}\left( \Delta z_{i,i-1} + \Delta z_{i+1,i} \right)}, \quad
a_{i,i+1}  = \frac{ s_i(t)}{ \Delta z_{i+1,i}\left( \Delta z_{i,i-1} + \Delta z_{i+1,i}  \right)},   \\
a_{ii} &= - a_{i,i-1} - a_{i,i+1}, \quad a_{i+j,i} = a_{i,i+j} = 0, \qquad j > 1.  \nonumber
\end{align}
\noindent where we defined a grid-valued set of {\it speed factors} (SFs)
\beq
\label{SF1}
s_i(t) = \hat{\sigma}^2(z_i,t).
\eeq
Now we have a Markov generator parametrized by the pointwise set of speed factors in \eqref{SF1}.
Next we will show how the SFs in \eqref{SF1} are turned into tunable parameters and used for
calibration to option quotes.

\subsection{Parametrization of Speed Factors in USLV(1,0)} \label{sect:N1_USLV_param}
As it stands, the parametrization in \eqref{empir_matrix} is very general. The number of free
parameters
per a given time slice $ t$ is $ p $, typically far exceeding the number of observed option
quotes available for
calibration for cases of practical interest.\footnote{Typical values of $ p $ that we have in mind
for practical implementation are around 20 to 100; see also \citet{CLS}.}

To achieve an exact match between the number of free parameters in our model and the number
of available option quotes, we consider the following parametrization of our 1D speed factors
\eqref{SF1}. Our approach here is similar to  how contagion factors are used in the BSLP model
of \citet{AH}.

We assume that as a function of time $ t $, the SFs $ s_i(t) $ are piecewise constant
between maturities of traded options. This considerably simplifies computation of matrix
exponentials.

For the dependence of $ s_i (t) $ on the grid index $ i $ (i.e., for the {\it $ z $ dependence}),
we proceed as follows. Let $K_0, K_1, \ldots, K_{q-1}$ be a set of strikes for traded
options across all maturities, expressed in terms of the $ Z$-space as, e.g.,
in Proposition 3 of \citet{CGW}.
We assume that all these strikes correspond to $ q $ different nodes on our grid.\footnote{This is
because our grid is constructed in such a way that it puts all strikes exactly at some nodes,
plus add some nodes in between and beyond the range of quoted strikes.}
We then model the speed factors $ s_i $ for all values of $ i $ by picking exactly $ q $ free
values $ \hat{s}_0, \ldots, \hat{s}_{q-1} $ at locations  $K_0, K_1, \ldots, K_{q-1}$,
and using linear interpolation for points in between. In other words, our speed factors $ s_i(t)$
are piecewise-linear in $ z_i $, while the anchor points at $ q $ selected nodes serve
as free parameters for calibration to option quotes.  Any consistent set of option quotes can be
exactly matched by the present method.\footnote{Alternatively, if the number of liquid option quotes
per maturity is large while the calibration speed is an issue, then some strikes can be
omitted from the set of anchor points at the price of giving up an exact calibration for these omitted
strikes, while keeping an exact calibration for the other strikes. For example, one can be guided by the
size of quoted bid-ask spreads in deciding which strikes could be skipped without much
sacrifice to accuracy while gaining in performance. To preserve no-arbitrage for the omitted
strikes, one should use a monotonic interpolation in the probability space.}

Note that calibration of local volatility models without assuming availability of a complete
set of option quotes  (i.e., when the number of option
quotes is less then the number of nodes on a grid)
 has been previously discussed in the literature. In particular, a recent
paper by \citet{Lipton_Sepp_2011} analyzes a setting where the local volatility function is
piecewise-flat between quoted strikes (or between mid-points) using Laplace-transform based
methods. Unlike their method, which is exact in 1D, our approach is based on numerical
optimization, but it is extendable to a multivariate setting (2D and higher)
along the same lines as in 1D. In addition, a piecewise-linear volatility model appears to be a less
drastic approximation than a piecewise-flat model.

\subsection{Calibration of the USLV(1,0) Model}
Calibration of the speed factors $ s_i(t) $  in the above setting is straightforward.
The anchor points introduced above serve as parameters of optimization.
Given a multidimensional optimizer, at each iteration
we first construct the generator matrix given the current set of the anchor points.
After that, finite-time probability distributions are computed by taking matrix
exponentials of the generator. As the mathematical structure of
our model is essentially the same for the $ N=1$  and $ N=2 $ cases (one or two
factors for the term structure), we postpone presenting
details of this procedure until the next section where we introduce an $ N=2 $ version of our model.
Theoretical option prices for a given set of model parameters are then computed using
these probability distributions.  Finally the optimizer adjusts the current set of free parameters to
decrease the error between the model and the market.

\section{USLV(2,0): Two Curve Factors} \label{sect:USLV_N2}

With two factors for the curve ($ N = 2 $), we assume the following
vector-valued SDE for the
dynamics of $ {\bf Z}_t = \left(Z_t^1,Z_t^2 \right)^\top $:
\bea
\label{Z_SDE_2D}
\left( \begin{array}{cc}
 d Z_t^{(1)} \\
d Z_t^{(2)}\\
\end{array}
\right)  =
\left( \begin{array}{cc}
 \Sigma_{11} &  \Sigma_{12}  \\
\Sigma_{21} & \Sigma_{22}  \\
\end{array}
\right)
\left( \begin{array}{cc}
 d W_t^{(1)}   \\
 d W_t^{(2)}  \\
\end{array}
\right),
\eea
where two Brownian motions $ W_t^{(1)}, W_t^{(2)} $ are independent, and the volatility matrix $ \Sigma = \Sigma(\bf z)$ is defined as follows:
\bea
\label{Sigma_2D}
\Sigma({\bf z}) =
\sqrt{s({\bf z},t)}
\left( \begin{array}{cc}
 \sigma_{1}\sqrt{1-\rho^2} &  \sigma_{1}\rho   \\
 0 & \sigma_{2}   \\
\end{array}
\right), \; \; \;  \Sigma({\bf z}) \Sigma({\bf z})^{T} =
s({\bf z},t)
\left( \begin{array}{cc}
 \sigma_{1}^2 &  \rho \sigma_1 \sigma_2  \\
 \rho \sigma_1 \sigma_2  & \sigma_2^2   \\
\end{array}
\right).
\eea
Here $ s({\bf z},t) = s({\bf Z}_t,t | {\bf Z}_t = {\bf z}) \geq 0 $ is a scalar function of the
state variable $ {\bf Z}_t =  \left(Z_t^{(1)}, Z_t^{(2)} \right)^\top $, and ${\bf z} = (z_1,z_2)$ are the values of ${\bf Z}_t$ at time $t$.
An explicit specification of this function will be given below. Note that
components $ Z_t^{(1)} $ and $ Z_t^{(2)} $
defined by \eqref{Z_SDE_2D} and \eqref{Sigma_2D} are correlated with correlation
coefficient $ \rho $.

Our first objective is to approximate the dynamics given by \eqref{Z_SDE_2D} by
a 2D Markov chain. To this end, we start with a Markov generator corresponding to
the 2D diffusion given by \eqref{Z_SDE_2D}:
\beq
\label{Markov_generator}
\mathcal{L} V( {\bf z}) = \frac{1}{2} \sum_{i,j = 1}^{N=2}
\left[ \Sigma \Sigma^{T} \right]_{ij} \frac{ \partial^2 V(\bf z)}{\partial z_i \partial z_j},
\eeq
where $ V({\bf z}) = V({\bf z}_T| {\bf z}) $ is an arbitrary function (a value function or
a transition density) of backward
 variables $ {\bf z}$ (with
forward variables ${\bf z}_T $ treated as parameters).
The generator specifies the continuous-space backward equation
\[
\frac{\partial V \left( {\bf z},t \right) }{\partial t} =
- \mathcal{L} V \left( {\bf z}, t  \right).
\]



Note that in order to be probabilistically interpretable as a generator of a Markov chain,
a discrete version $ A $ of the generator $ \mathcal{L} $ should have all nondiagonal elements
nonnegative, and all diagonal elements negative, such that all rows sum up to one.
These remarks are important as
not any discretization of $\mathcal{L}$ gives rise to
a valid Markov chain generator.
For example, using a central divided difference to approximate the mixed derivatives
in \eqref{Markov_generator} would not preserve nonnegativity of nondiagonal elements of $A$.

With these remarks in mind, and given a two-dimensional grid\footnote{For simplicity, in this section we assume that our one-dimensional
grids are uniform with the same number $p$ of grid points per each dimension.
Therefore, $z_{i+1,j} - z_{i,j} = \delta z_1, \ \forall j=1,N, \ i \in [1,p)$,
$z_{i,j+1} - z_{i,j} = \delta z_2, \ \forall i=1,N, \ j \in [1,p)$. For analysis of a nonuniform grid, see Appendix~\ref{ApNUG}.} of values of $ \left(Z^{(1)}, Z^{(2)} \right) $ with
$ p $ nodes per dimension,
we approximate second derivatives by divided differences. Derivatives $ V_{z_k,z_k}, \ k=1,2$, are represented using
the central differences
\begin{align*}
\left. \frac{\partial^2 V}{\partial z_1^2} \right|_{ij}  &=  \frac{V_{i+1,j} - 2
V_{i,j} + V_{i-1,j}}{(\Delta z_1)^2} + O \left((\Delta z_1)^2 \right), \\
\left. \frac{\partial^2 V}{\partial z_2^2} \right|_{ij}  &=  \frac{V_{i,j+1} - 2
V_{i,j} + V_{i,j-1}}{(\Delta z_2)^2} + O \left((\Delta z_2)^2 \right),
\end{align*}
\noindent while for the mixed derivative we take uncentered differences which preserve nonnegativity:
\begin{align*}
\left. \frac{\partial^2 V}{\partial z_1 \partial z_2} \right|_{ij} &=
\frac{V_{i+1,j+1} - V_{i+1,j} - (V_{i,j+1} - 2 V_{ij} +  V_{i,j-1}) - (V_{i-1,j} - V_{i-1,j-1})}{
2 \Delta z_1 \Delta z_2} \\
&+ O \left( \Delta z_1 \Delta z_2\right) + O \left( (\Delta z_1)^2\right) + O \left( (\Delta z_2)^2 \right), \quad  \rho \ge 0, \nonumber \\
\left. \frac{\partial^2 V}{\partial z_1 \partial z_2} \right|_{ij} &=
\frac{ V_{i+1,j} - V_{i+1,j-1} + (V_{i,j+1} - 2 V_{ij}  + V_{i,j-1}) -(V_{i-1,j+1} - V_{i-1,j}) }{
2 \Delta z_1 \Delta z_2} \\
&+ O \left( \Delta z_1 \Delta z_2\right) + O \left( (\Delta z_1)^2\right) + O \left( (\Delta z_2)^2 \right), \quad  \rho < 0.
\end{align*}
Using this in \eqref{Markov_generator} and regrouping terms, we obtain
\beq
\label{Markov_gen_2D}
\left( \mathcal{L} V(z) \right)_{ij} = \sum_{k,m \in \{ -1,0,1 \} } a_{ij|i+ k,j+m},
V_{i+k,j+m}
\eeq
where we introduced the following compact notation:
\begin{align}
\label{a_elements}
a_{ij | i+1,j} &= a_{ij | i-1,j} = \frac{1}{2} s_{ij}(t) \left( \frac{\sigma_1^2}{
\left( \Delta z_1 \right)^2} - \frac{ |\rho| \sigma_1 \sigma_2}{ \Delta z_1 \Delta z_2} \right),
\nonumber \\
a_{ij | i,j+1} &= a_{ij | i,j-1} = \frac{1}{2} s_{ij} (t) \left( \frac{\sigma_2^2}{
\left( \Delta z_2 \right)^2} - \frac{ |\rho| \sigma_1 \sigma_2}{ \Delta z_1 \Delta z_2} \right),
\nonumber \\
a_{ij| ij} &=  - s_{ij} (t)\left( \frac{\sigma_1^2}{
\left( \Delta z_1 \right)^2}  - \frac{ |\rho| \sigma_1 \sigma_2}{ \Delta z_1 \Delta z_2}
+ \frac{\sigma_2^2}{ \left( \Delta z_2 \right)^2}  \right), \\
a_{ij|i+1,j+1} &= a_{ij|i-1,j-1} = \left
\{ \begin{array}{ll}
\frac{s_{ij}(t)}{2} \frac{ \rho \sigma_1 \sigma_2}{ \Delta z_1 \Delta z_2 } & \mbox{if $ \rho \geq 0 $},  \\
0 & \mbox{if $ \rho < 0 $},
\end{array}
\right.  \nonumber \\
a_{ij|i+1,j-1} &= a_{ij|i-1,j+1} = \left\{
\begin{array}{ll}
  0 & \mbox{if $ \rho \geq 0 $},  \\
  \frac{s_{ij}(t)}{2} \frac{ |\rho| \sigma_1 \sigma_2}{ \Delta z_1 \Delta z_2 } & \mbox{if $ \rho < 0 $},
\end{array}
\right.  \nonumber
\end{align}
\noindent where $ s_{ij}(t) = \left[ s(Z_t,t) \right]_{ij} $.

To ensure that all parameters $ a_{ij|i+k,j+m}, \ k,m \ne 0 $ are nonnegative,
we impose the following constraint on the step size $ \Delta z_2 $
given a chosen step $ \Delta z_1 $:
\beq
\label{Delta_y_constraint}
| \rho | \frac{ \sigma_2}{\sigma_1} \Delta z_1 \leq \Delta z_2 \leq \frac{\sigma_2}{ | \rho| \sigma_1}
\Delta z_1.
\eeq
Assuming \eqref{Delta_y_constraint} is satisfied, we can interpret \eqref{Markov_gen_2D} as
a generator matrix of a 2D Markov chain. We can write it in a matrix form as follows:
\beq
\label{Markov_gen_discrete}
\left( \mathcal{L} V(z) \right)_{ij} = \left[ A V \right]_{ij} = \sum_{i',j'} A_{ij | i'j'} V_{i'j'}.
\eeq
As we deal with a two-factor setting, the matrix elements of the generator $ A $
carry four indices rather than two.
To sum over two indices corresponding to the
$ Z^{(1)}$- and $ Z^{(2)}$-states in \eqref{Markov_gen_discrete}, it is convenient
to group
all transitions according to the change of one variable (e.g., $ Z^{(1)} $). We obtain
\bea
\label{Markov_gen_2D_2}
&& \left( \mathcal{L} V(z) \right)_{ij} = \sum_{i',j'} A_{ij | i'j'} V_{i'j'}
=  \sum_{j'} A_{ij | i-1,j'} V_{i-1,j'} + \sum_{j'} A_{ij | ij'} V_{ij'} +
\sum_{j'} A_{ij | i+1,j'} V_{i+1,j'}
\nonumber \\
&& =
\left[ \left\{ a_{ij | i-1,j-1} V_{i-1,j-1} + a_{ij | i-1,j} V_{i-1,j} +
a_{ij| i-1,j+1} V_{i-1,j+1} \right\}
\right. \nonumber \\
&&\qquad \left. + \left\{a_{ij | i,j-1} V_{i,j-1} + a_{ij | ij} V_{ij} + a_{ij| i,j+1} V_{i,j+1} \right\}
\right.  \\
&&\qquad \left. + \left\{a_{ij | i+1,j-1} V_{i+1,j-1} + a_{ij | i+1,j} V_{i+1,j}
+ a_{ij| i+1,j+1} V_{i+1,j+1} \right\}
\right]. \nonumber
\eea
Here terms in the first, second, and third row correspond to transitions
$ i \rightarrow i - 1 $, $ i \rightarrow i $, and $ i \rightarrow i + 1 $ in the $ Z^{(1)}$-dimension,
respectively.

Mathematically, this is expressed via the following tridiagonal
block-matrix form for the resulting ``one-dimensional'' Markov chain generator $ A $:
\bea
\label{A_2D}
A \, = \, \left( \begin{array}{clcrccc}
L^{(0)}  & F^{(0)} & 0 & 0 & \cdots & 0  & 0  \\
B^{(1)} & L^{(1)} & F^{(1)}  &  0 & \cdots & 0 & 0 \\
0 &  B^{(2)}  & L^{(2)} & F^{(2)}  & \cdots & 0 & 0 \\
\vdots                        \\  	
 0  & 0 & 0 &  0 & \cdots & B^{( p-1 )}  & L^{( p-1 )}   \\
\end{array} \right), 
\eea
where all matrices $ B^{(i)}, L^{(i)},\, F^{(i)}$ have
dimension $ p\times p$, i.e.,
the dimension of our one-dimensional grids.\footnote{If grids in $ z^{(1)} $ and
$ z^{(2)} $ have different
lengths $ p_1 $ and $ p_2 $, then the size of these matrices will be $ p_2 \times p_2 $.}
Explicit expressions for these matrices
can be found from \eqref{Markov_gen_2D_2}:
\bea
\label{a_elements_2}
&& B_{j,j-1}^{(i)} = a_{ij | i-1,j-1}, \; B_{j,j}^{(i)} = a_{ij | i-1,j} , \;
B_{j,j+1}^{(i)} =  a_{ij | i-1,j+1}, \nonumber \\
&& L_{j,j-1}^{(i)} = a_{ij | i,j-1}, \; L_{j,j}^{(i)} = a_{ij | ij} , \;
L_{j,j+1}^{(i)} =  a_{ij| i,j+1,}, \\
&& F_{j,j-1}^{(i)} = a_{ij | i+1,j-1}, \; F_{j,j}^{(i)} = a_{ij | i+1,j} , \;
F_{j,j+1}^{(i)} =  a_{ij| i+1,j+1},  \nonumber
\eea
while all other elements of these matrices vanish. Note that this implies that the generator
\eqref{A_2D} is ``doubly'' sparse, as matrices $ B^{(i)}, L^{(i)} $ and
$ F^{(i)}$ are themselves
sparse; see also a comment at the end of this section.\ednote{End of the block to remove}



The block-tridiagonal matrix structure \eqref{A_2D} of the Markov chain generator $ A $
is characteristic of so-called quasi-birth-death (QBD) processes.
A QBD process is a bivariate Markov chain of a special type of dynamics of two components.
The first component,
$ Z_t^{(1)}$, called the ``level,'' follows a birth-and-death (BD) process on either a finite or infinite
set of states.
Conditional
on the realization of the $ Z_t^{(1)}$-component at a given step $[t,t+dt] $, the
second component $ Z_t^{(2)} $, called the ``phase,'' follows another Markov process.
For a short review of QBD processes, see, e.g., \cite{Kharoufeh2011}.
Note that in our particular case, $ Z_t^{(2)} $
follows another BD process,
while the support of $ Z_t^{(1)} $ is finite. QBD processes with finite support are called
finite QBD processes.



The symbols $L, B$ and $F$ in \eqref{A_2D} stand for local (without change of level),
backward and forward (the level is changed by one unit up or down) moves, respectively.
Note that as long as the matrices $ L^{(i)}, B^{(i)} $ and $ F^{(i)} $
depend on level $ i $ via the discretized local volatility function $ s_{ij} $, our
QBD process with generator \eqref{A_2D}
is a {\it level-dependent} (or {\it nonlinear}) finite QBD, denoted
sometimes as a (finite) LDQBD in the literature.

It is easy to check that \eqref{A_2D} with elements defined as in \eqref{a_elements}
is a valid Markov generator where all off-diagonal
elements are positive, diagonal elements are negative and
each row in $ A $ sums to zero.


After the QBD generator matrix is constructed according to \eqref{A_2D}, a matrix $ P$
of finite-time transition probabilities
with
matrix elements
\beq
\label{condForwardJoint}
P_{ij |i'j'} (t,T) \equiv
P \left[ (Z_T^{(1)},Z_T^{(2)} =  z_{i'}^{(1)}, z_{j'}^{(2)} |
Z_t^{(1)},Z_t^{(2)} =  z_{i}^{(1)}, z_{j}^{(2)} \right]
\eeq
can be computed by solving the forward Kolmogorov equation
\beq
\label{forward2D}
\frac{ \partial P}{ \partial T}  =   P A.
\eeq
For a given interval $ [t_1, t_2] $ where the generator $ A $ does not depend on time,
the solution of the forward equation is
\[
P = P_{t_1} e^{ (t_2 - t_1) A},
\]
where $ P_{t_1} $ is a state vector at time $ t_1$, and
$ e^{ X }$ stands for the matrix exponential of $ X $.

A few remarks on the complexity of the method just presented are in order here.
We have managed to map the two-factor continuous-space dynamics \eqref{Z_SDE_2D}
on the state space $ Z_1 \times Z_2 $
onto a QBD process with generator \eqref{A_2D}. The latter can formally be viewed as
a one-dimensional Markov chain in an extended linear space whose basis
is formed by elements of a Kroneker product of grid values
$ {\bf Z}_{g}^{(1)} \otimes {\bf Z}_{g}^{(2)} $ (and properly rearranged to form a QBD
structure). Therefore, computation of transition probabilities \eqref{forward2D} in our {\it two-factor} model is,
at least formally, as simple as a corresponding calculation for a {\it one-factor} model, and
reduces to computation of a single matrix exponential, albeit of a larger
matrix.\footnote{Here in addition to various efficient algorithms for computing
a matrix exponential of a sparse matrix, one could use
splitting in different dimensions that would take into account a block-diagonal form of
the generator matrix $A$.}

While naively the generator $ A $ has $ O(p^4) $ free parameters, their actual
number is much lower due to sparsity of the
matrix. It is simple to find that the number of nonzero elements that need to be stored
scales as $ (3p-2)p + 2(2p-1)^2$. For example, for $ p= 100 $ our matrix $ A $ would be of
size $ 10000\times10000 $ with only 109,002 non-zero elements. Matrices
of such sizes can well be handled by modern matrix exponentiation methods
(see below).\footnote{While a randomization method that we describe in the following section
may not be the most efficient method when the $L_2$ norm of matrix $A$ is
large (\cite{SS99}), it is very convenient for introducing stochastic volatility via
a stochastic time change. We therefore stick to this approach in what follows.}

\subsection{Transient Probabilities of QBD Chain by Randomization} \label{sect:randomization}
It is well known that a direct computation of a matrix exponential $ e^{t A} $ with a
Markov generator $ A $ via a straightforward use of a Taylor series expansion
as $ \sum_{n=0}^{\infty} (t A)^n/n! $ is in general not a good idea (see
\citet{Moler_2003}).
 The main reason for this is that severe roundoff errors might accumulate (especially
 when the matrix  is large) due to the fact that the generator has
 both positive and negative entries. In addition,
 matrices $ (t A)^n $ become nonsparse even if the original matrix $ A $ is sparse, as is the case
 for the QBD process.

 An efficient method of choice for dealing with matrix exponentials for large matrices is known
 as {\it Jensen's randomization}; see, e.g., \citet{GM}, or \citet{Haverkort} for a more recent
 review. The method proceeds as follows. We start
 with choosing a
 parameter $ \lambda  \geq \max_{i} \left\{ |A_{ii} | \right\} $, and define the matrix
 \beq
 \label{P_mat}
 {\bf P} = {\bf I} + \frac{{\bf A}}{\lambda} \; \Rightarrow {\bf A} =
\lambda \left( {\bf P} - {\bf I} \right).
 \eeq
 With our choice of $ \lambda $, all entries of $ {\bf P} $ are between 0 and 1, and all
 rows sum to 1. This means that $ {\bf P} $ is a stochastic matrix that describes a {\it
 discrete-time} Markov chain ``related'' to the original {\it continuous-time} Markov
 chain with generator $ A $.  We now want to discuss in more detail the sense in which these
 two Markov chains are ``related.''

 To this end, we substitute $ A $ as given by \eqref{P_mat} into the solution of the forward
 equation:
 \[
 P(t) = P_0' e^{t A} = P_0' e^{ \lambda t \left( {\bf P} - {\bf I} \right)} =
 P_0' e^{- \lambda t } e^{ \lambda {\bf P} t}.
 \]
 Using a Taylor series expansion for the matrix exponential in this expression, we obtain
 \beq
 \label{Taylor_P}
 P(t) = P_0' e^{- \lambda t } \sum_{n=0}^{\infty} \frac{ \left( \lambda t \right)^n {\bf P}^n}{n!}
 = P_0'  \sum_{n=0}^{\infty} \psi( \lambda t, n) {\bf P}^n,
 \eeq
 where
 \[
 \psi(\lambda t, n) = e^{ - \lambda t} \frac{ ( \lambda t )^n}{n!} , \; \; n \in \mathbb{N},
 \]
 are Poisson probabilities, i.e., probabilities of observing $ n $ events by time $ t $ for
 a Poisson process with intensity $ \lambda $. Note that a naive Taylor expansion
 of the matrix exponential $ e^{ t A} $ behaves badly, but the new expansion is much better
 behaved: roundoff errors are now
 largely eliminated as all entries of matrix $ {\bf P} $ are between 0 and 1. Moreover, different
 terms are weighted by the Poisson probabilities, so that the expansion is expected to converge
 fast when the product $ \lambda t $ is not too large.

 We note that the construction given by \eqref{Taylor_P} can be interpreted as a
 discrete-time Markov
 chain (DTMC) $ Y_n $ ($n = 0,1,\ldots,p $) with transition matrix $ {\bf P} $
 subordinated to a Poisson process $ N_t $ where the latter serves as a randomized ``operational
 time'' for $ Y_n $, so that the  subordinated process is now defined as
 $ X_t = Y_{N_t} $ (see, e.g., \citet{Feller}). We will return to the topic of
 subordinated processes in Sect.~\ref{sect:stoch_time_changes}.

It is important to point out that the method just presented
can be used without a matrix-matrix
multiplication (as \eqref{Taylor_P} would naively suggest). Let $ {\hat {\bf \pi}}_n $ be
the probability
distribution vector in the DTMC with transition matrix $ {\bf P} $ after $ n $ epochs.
This vector can be computed recursively:
\bea
\label{recursive_PP}
&& {\hat {\bf \pi}}_0 = {\bf P}_0,  \nonumber \\
&& {\hat {\bf \pi}}_n = {\hat {\bf \pi}}_{n-1} {\bf P} , \; \; n \in {\mathbb N}^{+}.
\eea
Using the vector $  {\hat {\bf \pi}}_n $, the state probabilities \eqref{Taylor_P} in the original CTMC
are computed as follows:
\[
 P(t)
 =  \sum_{n=0}^{\infty} \psi( \lambda t, n)  \left( P_0'  {\bf P}^n \right) =
 \sum_{n=0}^{\infty} \psi( \lambda t, n) {\hat {\bf \pi}}_n.
 \]
Therefore, computationally the algorithm amounts to a series of vector-matrix multiplications
that can be done very efficiently for matrices of sizes typical for our problem.
Moreover, the recursive procedure of \eqref{recursive_PP} preserves the sparsity
of $ {\bf P} $, thus enabling a substantial acceleration of the vector-matrix multiplication.

In practice, the infinite sum in \eqref{Taylor_P} is truncated at some value $ n_{max} $. This value can be adaptively controlled within the algorithm
itself, as a theoretical upper bound for an error resulting from the truncation is available as
discussed, e.g., by Gross and Miller (1984).

\subsection{Calibration of Speed Factors in USLV(2,0)}
\label{calib_N2}

Summarizing our results for the USLV(2,0) model so far, we see that the mathematical
structure of the (2,0) model is similar to that of the (1,0) model. Indeed, for the latter
our Markov chain construction gives rise to a {\it nonlinear birth-death} (BD) process
modulated by 1D speed factors (SFs) $ s_i(t) $. After a proper parametrization as described
in Sect.~\ref{sect:N1_USLV_param}, SFs are calibrated to market option quotes. For the
(2,0) case (two factors for the term structure), the resulting Markov chain is a {\it
nonlinear quasi-birth-death} process, again modulated by a set of SFs $ s_{ij} $.
In terms of computational complexity, the two cases are essentially
the same, as both
amount to calculation of matrix exponentials of a Markov chain generator (albeit in different
dimensions).

Calibration of the (2,0) model is done by the fitting function $ s(z,t) $ in \eqref{Z_SDE_2D}.
To this end, we proceed similarly to the one-factor case. We assume that
function $ s(z,t) = s(z_1, z_2,t) $ is a function of single argument, $ s(z_1,z_2,t) =
s(\alpha_1 z_1 + \alpha_2 z_2,t) $, where $ \alpha_1, \alpha_2 \geq 0 $ are some weights (e.g., $ \alpha_1 = \alpha_2 = 0.5 $).
%
For calibration purposes,
we could parametrize this function in a piecewise-linear way, i.e., in exactly the same way as
we did before for the $ N = 1 $ model. The number of free parameters (anchor points) and
their locations would be chosen based on a particular set of instruments available for calibration.

To continue with our theoretical construction of the model, in what follows we assume that
the stage of construction of a (2,0) (or (1,0)) version of the USLV model is
completed along the lines described here. In what follows, we refer to these SFs as
1D SFs, in order to differentiate them from another set of speed factors (2D SFs)
that will be introduced below when we add stochastic volatility to the model.

Finally, we note that while the main purpose of the USLV(2,0) model
for our purposes is to use it as a building block in the construction of a full-blown
USLV(2,2) model with stochastic volatility, the pure local volatility USLV(2,0)
model can also be useful in its own right, e.g., as a way of pricing European vanilla
options with illiquid strikes in terms of prices of liquidly traded options.

\section{USLV(2,2): Two Curves and Two Volatility Factors}
\label{sect:USLV_N2_M2}


Once we have a calibrated USLV(2,0) model, introduction of stochastic volatility
in this framework amounts to two things: (i) introducing new dynamics for volatility drivers, and
(ii) making sure the model still calibrates to available option prices. This produces
a calibrated USLV(2,2) model.

Let us note that stochastic volatility dynamics can be introduced in our framework in two
ways. In the first approach, we follow  the formulation of a continuous-time Markov
chain (CTMC) dynamics, which we now augment by 2D dynamics of ``spot'' variance factors.
For numerical implementation, the model is then put on a time grid
$[t_0, t_1, \ldots, t_n] $ with a uniform time step $ \Delta t $. All calculations (see below)
are done to $ O( \Delta t^2) $ accuracy, which assumes that $ \Delta t $ should be
sufficiently small.\footnote{For example, we might need to use daily or more frequent steps, depending
on the level of volatility, with this approach.}  With this method, we solve the forward and backward
Kolmogorov equations one step $ \Delta t $ at a time, similarly to finite difference methods.

In the second approach, we deal with arbitrary time lines which do
not necessarily
have small time steps. For example, we may want to model the values of underlying factors
only on a sparse set of ``interesting'' dates (e.g., coupon dates, call dates etc.).
Essentially, by taking matrix exponentials of the generator, we aim here to achieve a functionality similar to the USLV(1,0)  and USLV(2,0) models
(or any CTMC model, for that matter), which are capable of computing
finite-time transition
probabilities directly.\footnote{To the extent that
one-step methods, such as the Runge-Kutta method, can be viewed as particular ways
to compute matrix exponentials (see \citet{Moler_2003}), what we mean here by ``direct''
calculations are
other methods of computing matrix exponentials that might in some cases
be more efficient than one-step
methods.}

Respectively, in what follows we present two versions of the USLV(2,2) model.
In the first version, we assume a Markov dynamics in the pair $ \left( {\bf Z}_t, {\bf Y}_t \right)$
where $ {\bf Y}_t = \left( Y_t^{(1)}, Y_t^{(2)} \right) $ is a bivariate ``spot'' variance
driver.
In the second version, we instead assume a Markov dynamics in a pair of $ {\bf Z}_t $ and
an {\it integrated} bivariate variance, or, more generally, a
bivariate stochastic time subordinator (see below).
We will use the notation $ {\bf T}_t = \left( T_t^{(1)}, T_t^{(2)} \right)$ for the latter in what follows.
For reasons that will become clear shortly, we will refer to these two versions of the model
as the {\it activity-rate}-based model (AR-USLV), or the {\it implied-time-change}-based model
(ITC-USLV),  respectively.

On the {\it theoretical} side, it turns out that both approaches can be viewed in a unified way
by interpreting them as particular realizations of a stochastic time change of the original QBD
Markov chain. We will give more details on this below in Sect.~\ref{sect:stoch_time_changes}.

On the {\it practical} side, we can choose between two numerical methods.
With the first method, we can implement both the
AR- and ITC-versions of our model in a similar
way using a version of the Markovian projection method.  The latter reduces calibration of
USLV(2,2) to a fast forward induction method in what is essentially
a 1D problem, without a need for a forward induction on a
full 2D Markov chain.\footnote{Recall that by 1D and 2D, we mean
linearized spaces obtained from the pairs $ \left( Z_t^{(1)}, Z_t^{(2)} \right) $ and
$ \left( Y_t^{(1)}, Y_t^{(2)} \right) $ by taking elements of pairwise Kroneker products
as new 1D
bases. In terms of factor counting, our 1D and 2D Markov chains correspond to the two- and
four-factor model specifications, respectively.} No new optimization in addition to one
performed at the stage
of calibration of the USLV(2,0) model is involved here. Therefore, the method is very fast
on each given time step, the only potential bottleneck being the necessity to perform such
computation repeatedly on a dense time grid. The method is {\it nonparametric}
in that it solves the
problem of calibration of the full-blown USLV(2,2) model via a judicious choice of 2D
speed factors (SFs) that are computed off the calibrated 1D SFs of the USLV(2,0) model.

The second method, which is applied below for the ITC-USLV version of our model
but in principle could be used for both versions,  is to
``break the symmetry,'' and make the process $ {\bf T}_t = \left( T_t^{(1)}, T_t^{(2)} \right) $
{\it parametric} in one dimension (e.g., $ T^{(2)} $), while keeping it {\it nonparametric}
in another dimension (resp., $ T^{(1)} $). The idea here is that for the purpose of
calculation
of finite-time transition probabilities in the $ {\bf Z}$-space, we can perform
averaging over the
randomness due to $ T_t^{(2)} $ {\it analytically} (or semi-analytically) once a tractable
model
for subordinator $ T_t^{(2)} $ is specified. The averaging over the residual randomness due
to $ T_t^{(1)} $ is performed numerically. Similarly to the previous case, this
calculation can be
done in a nonparametric setting, where at each step on our sparse time grid, we introduce
just enough free parameters to match observed quotes for options maturing
at this time. {\it Differently} from the previous case, the recalibration
to option quotes in the present setting
amounts to a  (convex) optimization problem in the dimension equal to the number of
option quotes for this maturity.

The two flavors (AR and ITC) of our USLV(2,2) model outlined above
thus offer a certain trade-off in terms of complexity.
For the AR-USLV(2,2) model, the recalibration is fast for one step, but complexity
scales linearly with the number $ N_d $ of
time steps on a dense grid; i.e., the complexity is $O( N_d) $.  For the ITC-USLV(2,2) model, the
complexity is
$ O (N_s  N_c) $, where $ N_s $ is the number of nodes on a sparse
time line, and $ N_c $ is the number of option quotes per node, independently of  $ \Delta t $,
but the $ O(N_c) $ part above involves convex optimization in dimension $ N_c $.
Based on previous experience
with similar models, we expect a compatible performance from the two versions of the USLV(2,2)
model, at least for typical cases (e.g., $ N_c  = 5$, $ N_t = 40 $). Therefore, in what
follows we
will present both versions of the model.

Our plan for the reminder of this paper is as follows.
In the rest of this section, we describe the AR-USLV(2,2) version of the model,
where the Markov pair is  $ \left( {\bf Z}_t, {\bf Y}_t \right)$,
with a bivariate spot variance driver $ {\bf Y}_t = \left( Y_t^{(1)}, Y_t^{(2)} \right) $.
In Sect.~\ref{sect:BJN_overview},
we provide  a qualitative overview of this version of the model. The following subsections
of Sect.~\ref{sect:USLV_N2_M2}
provide details of our approach.
The ITC-USLV(2,2) version of the model, where the Markov pair is
$ \left( {\bf Z}_t, {\bf T}_t \right)$
with  $ {\bf T}_t = \left( T_t^{(1)}, T_t^{(2)} \right) $ being a bivariate subordinator, is
presented in Sect.~\ref{sect:USLV22_with_ITC}. As will be shown below, calibration to
observed option prices amounts, in this approach, to a construction of
an {\it implied time change}
(ITC) process. Within a particular approach  presented in  Sect.~\ref{sect:USLV22_with_ITC}, the
ITC process is defined in terms of  a bivariate exponential-L{\'e}vy process $ {\bf L}_t =
\left( T_t , \theta_t \right) $ where $ T_t $ is a {\it parametric} subordinator (e.g.,
an exponential gamma
process), and $ \theta_t $ is a {\it nonparametric} subordinator. The
latter will be
referred to as a
{\it time dilaton} process, for reasons explained below.

\subsection{Overview of AR-USLV(2,2)}
\label{sect:BJN_overview}

As was mentioned above, a model obtained from our USLV(2,0) model by
adding new state variables (in this case, spot volatility drivers
$ {\bf Y}_t = \left( Y_t^{(1)}, Y_t^{(2)} \right) $) would not in general
match observed option prices, even if our initial USLV(2,0) model does. Moreover, for
any particular parametric
model for the dynamics of the pair  $ \left( Y_t^{(1)}, Y_t^{(2)} \right) $, we are
still not guaranteed
that the full model could accurately fit available option quotes even after
calibration of parameters of the $ Y $-process.

In order to reinforce a nearly exact calibration to options for all consistent sets of
quotes, we
introduce
2D speed factors (SFs) $ S(z,y,t) $ in the full (2,2) model, that play an analogous role
to 1D SFs \eqref{SF1} in the (2,0) version of the model. We then provide a fast scheme to
compute 2D SFs based on solving the forward equation on the Markov chain. Our method
is similar to one used by \citet{BJN} (BJN), see also \citet{Rossi}, for
a lattice-based
stochastic local equity model in a (1+1) setting. A similar method was used
in the BSLP model by \citet{AH} for modeling dynamics of credit portfolios. For a similar
method used for equity option pricing, see \citet{Madan}.

A peculiar feature of a BJN-like forward induction method (to be presented in detail below)
is that it tries to adjust the $ Z $-process for {\it any} $Y$-process. It does not
address the problem of calibration of parameters of the $ Y $-process itself. In certain
situations, it might make sense to try to calibrate parameters of the $ Y$-process
as accurately as we can {\it before} adding a local volatility layer (so that our change to a
parametric model due to introduction of a local volatility would be a minimal tweak of
the model).
Alternatively, we could try to fit parameters of the $ Y$-process {\it after} we
introduce the
local volatility layers, but {\it before} we compute the 2D SFs.
This might be
an attractive option for a practical method of model calibration in our setting.
The reason is that if such parametric calibration of the $ Y $-process produces a good but
not perfect fit to the data, the role of non-parametric 2D SFs of the
$(N=2,M=2) $ model would be to perfect quality of calibration at the price of
adding some nonparametricity.\footnote{We hold a view that nonparametricity is ``evil,'' but
it is a ``common evil'' in the sense that it is used everywhere (for term structure
calibration, local
volatility models etc.).}

Note that while the BJN-like approach does not by itself address the problem of
calibration of
parameters of the $ Y$-process for a parametric specification of the dynamics of
$ {\bf Y}_t $,
this is where we could use Laplace transform based methods for stochastic subordinators,
similar to the method presented in Sect.~\ref{sect:USLV22_with_ITC} in a slightly different
setting. This implies that the calibration method presented later in this section and a method
presented in Sect.~\ref{sect:USLV22_with_ITC} can in practice be used together for a joint
parametric/nonparametric calibration of the AR-USLV(2,2) model.

Yet another possible way to calibrate our model would be as follows.
If a trader has a strong view on the relative weights of a spanned (delta-driven)
 and unspanned (genuine vega)
contribution to options' vegas, and wants the model to behave accordingly, this could
be achieved as follows.  Assuming that we are able to map constraints like those onto
some typical behavior of the set of SFs,\footnote{Such dependence can be established
either theoretically, or empirically on the basis of behavior of the model as a function of
model parameters.} we first fix some set of 1D SFs, and then calibrate
parameters of the $ Y$-process
given these SFs. After parameters of the $ Y$-process are specified in this way, we proceed
in the regular way of calibrating the model, by first computing the ``true'' (market-implied) set
of 1D SFs, and then following the forward calibration of 2D SFs in the full-blown model with
parameters of the $ Y$-process just computed at the previous step. Again, a combination
of various methods presented below can be used to implement such a calibration strategy.

As a brief summary, our QBD Markov chain stochastic-local volatility offers substantial
flexibility in how the model can be calibrated to available market and/or historical
data. Different steps/versions
of the calibration procedure can be combined (or skipped), depending on the
specific needs of an end user.
We now proceed with
describing our framework.

\subsection{QBD Processes and Stochastic Time Changes}
\label{sect:stoch_vol}

We prefer to think of stochastic volatility in terms of a stochastic
time change of some ``base'' process such as a Brownian motion. (See Sect.~\ref{sect:USLV22_with_ITC} for more details and relevant
references.) As our original two-factor $ (N=2) $ diffusion equation, \eqref{Z_SDE_2D}, has two Brownian
drivers, $ W_t^{(1)} $ and $ W_t^{(2)} $, we can use two {\it different} stochastic clocks
on them. This would amount to having a stochastic local volatility model
with $ N = 2 $ and $ M = 2 $.

Such formulation can be useful for asset classes where the short- and long-term
option volatilities typically behave differently (e.g., have different typical
levels or vol-of-vol),
in addition to a different behavior of
$ Z_t $-factors driving short- and long-term prices of basic instruments (bonds, futures etc.).
For example, for modeling commodity derivatives, we might want to have one
long-term factor $ Z_t^{(1)} $
driven by a Brownian driver $ W_t^{(1)} $ with its own
stochastic clock (stochastic volatility) driver $ Y_{t}^{(1)} $, and another, short-term factor
$ Z_t^{(2)} $ driven by another (possibly correlated)
Brownian driver $ W_t^{(2)} $, with its own stochastic clock
driver $ Y_{t}^{(2)} $.\footnote{Alternatively, correlated dynamics of two stochastic
drivers with each one
having its own stochastic volatility factor can be used for pricing hybrid derivatives.} This
results in a four-factor scenario with correlated long- and short-term factors, each
having its own stochastic volatility driver.

The above picture of two curve drivers each having its own stochastic clock is not
lost in our discrete-space Markov chain construction.
As we will show next, the structure of our QBD Markov chain
\eqref{A_2D} for the $ N= 2 $ case enables introducing two stochastic clocks
in the model in an internally consistent way, and without any need of introducing
additional {\it ad hoc} constraints on the model dynamics.
These stochastic clocks will modulate two Markov
chain generators. As the latter play the role of stochastic drivers in the discrete-space
setting, the resulting ``ecosystem'' of (discrete-valued) curve and volatility factors
bears a strong structural similarity to its continuous-space counterpart.

To explain our construction, we start with representing
the Markov generator \eqref{A_2D} in the following form:
\bea
\label{A_2D_two}
A & = & \left( \begin{array}{cccccc}
- \hat{F}^{(0)} & F^{(0)}           & 0            &   \cdots        & 0  \\
B^{(1)}            & - \hat{F}^{(1)} & F^{(1)}    &    \cdots            & 0 \\
0                     &  B^{(2)}          &  - \hat{F}^{(2)}   &    \cdots             & 0 \\
\vdots                        \\  	
 0                    & 0                    & \cdots    & B^{( p )}     & - \hat{F}^{( p )}
\end{array} \right) 
+
\left( \begin{array}{cccccc}
\hat{L}^{(0)}    & 0                   &  0           &    \cdots     & 0  \\
0                     & \hat{L}^{(1)}  &  0           &    \cdots     & 0 \\
0                     & 0                   &  \hat{L}^{(2)}    &    \cdots     & 0 \\
\vdots                        \\  	
 0                    &  0                   & 0            &   \cdots        & \hat{L}^{( p )}   \\

\end{array} \right)  \nonumber \\
& \equiv  & A_1 + A_{2},
\eea
where $  \hat{F}^{(i)} = \mathrm{diag} \left(
F^{(i)} {\bf 1} \right) +
\mathrm{diag} \left(B^{(i)} {\bf 1} \right) $ and
$ \hat{L}^{(i)} = L^{(i)} + \hat{F}^{(i)}  $, with $ {\bf 1} $ being a vector of ones.
Using \eqref{a_elements} and \eqref{a_elements_2},
it can be readily checked that both $ A_1 $ and $ A_{2} $
defined in \eqref{A_2D_two} are valid generators in the sense that for both,
all off-diagonal elements
are positive, all diagonal elements are negative and all rows sum up to zero.

This can be interpreted as follows. The second generator $ A_2 $ corresponds
to an idiosyncratic component of the $ Z$-dynamics
that is
independent of the rest of the system, and can be thought of as describing
idiosyncratic jumps of $ Z_t^{(2)} $ that occur
without a simultaneous jump of $ Z_t^{(1)} $ on the same time interval.\footnote{This can
also be viewed as a
one-dimensional orthogonal projection of two-dimensional dynamics of the pair $(Z_t^{(1)},
Z_t^{(2)}) $ onto a subspace where no jumps in variable $ X = Z_t^{(1)} $ are allowed.}
In terms of representation of stochastic dynamics of our system, this generator is an avatar
of
the idiosyncratic Brownian driver $ W_t^{(2)} $ in \eqref{Z_SDE_2D}. The first
generator $ A_{1} $
describes joint jumps of $ Z_t^{(1)} $ and $ Z_t^{(2)} $, and can be thought of as an avatar
of the
common Brownian driver $ W_t^{(1)} $ in \eqref{Z_SDE_2D}.

As shown in Appendix~\ref{AppA}, a random time change
of a continuous-time Markov chain amounts, in terms of the resulting Markov generator for the
chain, to
scaling all elements of the original Markov chain generator by a common factor given
by the value of
the activity rate (intensity of the time change) process.

As they conserve probability separately from each other, two generators
$ A_1 $ and $ A_{2} $ can be seen as representing two different subsystems of
our dynamic system in the $ (Z_t^{(1)},Z_t^{(2)}) $-space. As a time change acts as a common
scaling factor on a generator matrix (see Appendix B), this implies that
we can subject two generators, $ A_{1} $ and $ A_{2} $,
to {\it different} stochastic clock changes without any problems with laws of probability:
after separate time changes, probabilities are still all nonnegative, and sum up to one
in each subsystem separately. This gives rise to a discrete-space version of a continuous
$ M = 2 $ stochastic volatility model.

To summarize, the (2+2)-factor structure of the original continuous-space system
\eqref{Z_SDE_2D} (i.e. two factors for the term structure and two factors for the volatility)
is now naturally mapped onto a corresponding structure in our Markov chain model,
where a QBD process is an avatar of a two-dimensional Brownian motion
$ \left(W_t^{1)}, W_t^{(2)} \right) $, and two volatility factors are separately introduced as
two stochastic clocks for two generators $ A_1 $ and $A_{2} $ as explained above.
We now discuss specific realizations of this scenario in our model.

\subsection{ Forward Equation and Transition Probabilities in USLV(2,2)}


We assume discrete dynamics of stochastic intensity (stochastic volatility) drivers
$ {\bf Y}_t $, with an odd number
$ N_y = 2q + 1 $ of
discrete states
for each driver $ Y_t^{(1)}, Y_t^{(2)} $. Points on a
two-dimensional $ Y $-grid are denoted as
$ \left\{ Y_{\alpha_1}^{(1)}, Y_{\alpha_2}^{(2)} \right\}_{\alpha_1, \alpha_2 =
0}^{\alpha_1, \alpha_2 = 2q}$.  The initial values
$ (Y_{t=0}^{(1)}, Y_{t=0}^{(2)}) $ correspond to the midpoints $ (Y_{q}^{(1)},
Y_{q}^{(2)} ) $ on the grid.

In practice, we prefer to keep a low number of states (say 3 to 11) per volatility factor.
As volatility is unobservable, we feel that maintaining a low number of states might suffice to reproduce most
important stylized facts about stochastic volatility such as mean reversion and/or volatility
clustering (persistence), alongside its role in providing a better behavior of
a forward smile (a non-flattening smile for longer maturities) than typical behavioral
patterns observed
with local volatility models.

To ease the notation, in this section we use Latin indices $ i, j, k $ to enumerate
states ($ {\bf Z}_t = {\bf Z}_i, {\bf Z}_{t+dt} = {\bf Z}_j $ etc.), and Greek indices $ \alpha, \beta $
to enumerate
values of $ {\bf Y}_t $, $ {\bf Y}_{t+dt} $. However, because we deal with a two-factor
setting, both the indices and factor values are now two-component vectors rather than scalars;
for example,
\beq
\label{Z_component}
{\bf Z}_i  = \left( Z_{i_1}^{(1)}, Z_{i_2}^{(2)} \right)\, , \; \; i = ( i_1, i_2)\, , \; i_1, i_2 \in \mathbb{Z}^{+}.
\eeq
A similar representation is used for volatility drivers $ {\bf Y}_{\alpha} =
\left( Y_{\alpha_1}^{(1)}, Y_{\alpha_2}^{(2)} \right) $. In what follows we use both the vectorized
and component notations.

We postulate that the 2D dynamics of the pair $({\bf Z}_t, {\bf Y}_t)$ (where
both factors $ {\bf Z}_t $ and $ {\bf Y}_t $ are two-dimensional) in the USLV
model is Markovian.
The
system is defined in terms of the joint marginal probabilities
\[
\pi(j,\alpha,t) \equiv P \left[ {\bf Z}_t =  {\bf Z}_j, {\bf Y}_t = {\bf Y}_{\alpha} \right]
\]
and conditional transition probabilities
\[
 p_{i \alpha | j  \beta } (t, t + dt)
 \equiv P \left[ {\bf Z}_{t + d t}
= {\bf Z}_j, {\bf Y}_{t + dt} = {\bf Y}_{\beta} | {\bf Z}_{t} = {\bf Z}_i, {\bf Y}_{t} = {\bf Y}_{\alpha} \right].
\]
The forward equation takes the form
\beq
\label{forward}
\pi(j,\beta,t + dt) = \sum_{j,\alpha}
 p_{i \alpha | j  \beta } (t, t + dt) \pi(i ,\alpha,t).
\eeq
The transition probabilities have the following expansion:
\beq
\label{trans_prob_expansion}
p_{i \alpha | j  \beta } (t, t + dt) = \delta_{i j}  \delta_{\alpha \beta}
+ A_{i  \alpha | j  \beta} (t) dt + O \left( d t ^2 \right),
\eeq
where $ A_{i  \alpha | j  \beta} (t) $ is the Markov generator,
and $ \delta_{ij} = \delta_{i_1 j_1} \delta_{i_2 j_2} $ is the 2D Kroneker symbol
(with a similar definition for $ \delta_{\alpha \beta} $).

To proceed, we introduce the following conditional probabilities:
\bea
\label{cond_probs}
P_{i  j }^{(\alpha)}(t,t+dt) &=&  P \left[ {\bf Z}_{t + d t}
= {\bf Z}_j | {\bf Z}_{t} = {\bf Z}_i, {\bf Y}_{t} = {\bf Y}_{\alpha} \right],  \\
\hat{P}_{\alpha  \beta}^{(i j)} (t, t + dt) &=&
P \left[ {\bf Y}_{t + dt} = {\bf Y}_{\beta} | {\bf Y}_{t} = {\bf Y}_{\alpha},
{\bf Z}_{t} = {\bf Z}_i,  {\bf Z}_{t + d t}
= {\bf Z}_j  \right]. \nonumber
\eea
The joint probability $ p_{i \alpha | j \beta} $ can now be written as follows:
\beq
\label{composition_law}
p_{i \alpha | j  \beta } (t, t + dt) = P_{i  j}^{(\alpha)} (t, t + dt)
\hat{P}_{\alpha  \beta}^{(ij)} (t, t + dt).
\eeq
Using \eqref{trans_prob_expansion}, we obtain
\beq
\label{pAlpha}
P_{i  j }^{(\alpha)}(t,t+dt)  = \sum_{\beta} p_{i \alpha | j  \beta } (t, t + dt)
\equiv \delta_{ij} + \hat{A}_{i  j}^{( \alpha)} (t) dt + O \left( d t ^2 \right),
\eeq
where
\beq
\label{cond_MC_gen}
\hat{A}_{i  j}^{( \alpha)} (t) = \sum_{\beta} A_{ i \alpha | j \beta} (t) \, , \; \;
\sum_{j} \hat{A}_{i  j}^{( \alpha)} (t) = 0
\eeq
is the conditional generator of the $ Z$-Markov chain.

It  is convenient to write the second
conditional probability in \eqref{cond_probs} in the following form:
\bea
\label{cond_gen}
\hat{P}_{\alpha  \beta}^{(ii)} (t, t + dt) &=& \delta_{\alpha \beta}
+ \hat{Q}_{\alpha  \beta}^{(i)}(t) dt \, , \; \; \; \sum_{\beta} \hat{Q}_{\alpha  \beta}^{(i)} (t) = 0, \\
\hat{P}_{\alpha  \beta}^{(i,i+m)} (t, t + dt)
 &=& \delta_{\alpha \beta}  + \tilde{Q}_{\alpha  \beta}^{(m)}(t) \, , \; \;
 \sum_{\beta} \tilde{Q}_{\alpha  \beta}^{(m)} (t) = 0\, , \; \;  (m_1, m_2) \neq (0,0).
 \nonumber
 \eea
Note that the term $ \tilde{Q}_{\alpha  \beta}^{(m)}(t) $ in the second equation is {\it not} multiplied
by $ dt $ because the second relation in \eqref{cond_probs} is not a transition probability but
rather a conditional probability where we condition, in particular, on $ {\bf Z}_{t + dt} $.
If $ d {\bf Z}_t \neq 0 $, then $ dt $ cancels out in the calculation of the conditional
probability. This means that $ \tilde{Q}_{\alpha  \beta}^{(m)}(t) $ is not a
real generator, but rather a ``pseudo generator'' introduced here to simplify formulae to follow.
In its turn, this means that diagonal elements of $ \tilde{Q}_{\alpha  \beta}^{(m)}(t) $ cannot be made arbitrarily negative, as otherwise we would
end up with probabilities reaching outside of the unit interval $ [0,1] $.

Now we plug
\eqref{trans_prob_expansion} and \eqref{pAlpha} into \eqref{composition_law}.
This produces the following relation (here we omit $ O \left( dt^2 \right) $ terms):
\beq
\label{AaHat}
\delta_{i j}  \delta_{\alpha \beta}
+ A_{i  \alpha | j  \beta} (t) dt = \hat{P}_{\alpha  \beta}^{(ij)} (t, t + dt)
\left[ \delta_{ij} + \hat{A}_{i  j}^{( \alpha)} (t) dt \right].
\eeq
A more explicit expression for the generator $ A_{i \alpha | j \beta} $ in terms
of auxiliary generators $ \hat{A}_{i  j}^{( \alpha)} $,  $ \hat{Q}_{\alpha  \beta}^{(i)} $
and $ \tilde{Q}_{\alpha  \beta}^{(m)} $ can be obtained using the following identity
(which can be checked by inspection):
\beq
\label{identity}
A_{i \alpha | j \beta} = ( 1 - \delta_{ij}) A_{i \alpha | j \beta} +
(1 - \delta_{ \alpha \beta} ) A_{i \alpha | j \beta} - ( 1 - \delta_{ij})
(1 - \delta_{ \alpha \beta} ) A_{i \alpha | j \beta} +
\delta_{ij} \delta_{\alpha \beta} A_{ i \alpha | i \alpha}.
\eeq
Using \eqref{AaHat} to evaluate different terms in the right-hand side of \eqref{identity}
in terms of the auxiliary generators, we obtain, after some algebra, the following
general representation of generator $ A_{i  \alpha | j  \beta} (t) $  of USLV(2,2):
\beq
\label{Markov_gen_2_2}
A_{i \alpha | j \beta} = \delta_{ij} \hat{Q}_{\alpha  \beta}^{(i)} +
( 1 - \delta_{ij}) \tilde{Q}_{\alpha \beta}^{(j-i)}
\hat{A}_{ij}^{(\alpha)} +
\delta_{\alpha \beta} \hat{A}_{ij}^{(\alpha)}.
\eeq
Different terms in this expression are interpreted as follows.\footnote{We thank
Leonid Malyshkin for proposing a decomposition of the Markov generator in such form,
as well as
for discussions that helped improve the presentation in this section.}

The first term $
\delta_{ij} \hat{Q}_{\alpha  \beta}^{(i)}  $ is a generator of idiosyncratic
jumps of $ {\bf Y}_t $ that proceed without a simultaneous
jump of $ {\bf Z}_t $ in the interval $ [t, t + dt] $.
Various continuous-space models can be used as a means to parametrize this generator
via discretization of the state space. For example, starting with a diffusive model
for $ {\bf Y}_t $, we end up with a tridiagonal generator matrix
$ \hat{Q}_{\alpha  \beta}^{(i)}  $. More details and examples will be
given below in Sect.~\ref{sect:Discrete_Y_dynamics}.

 The second term  in \eqref{Markov_gen_2_2}
 is a generator of joint jumps of $ \left( {\bf Z}_t, {\bf Y}_t \right) $.
 Note that it is a valid generator on its own as long as
 $ \sum_{\beta} \tilde{Q}_{\alpha \beta}^{(m)}  = 0 $.
 Again, different specifications of this
 generator can be considered within our
 general framework. This will be discussed in some detail below in
Sect.~\ref{sect:Joint_jump}.

 Finally, the last term in  \eqref{Markov_gen_2_2} is a generator of
 idiosyncratic
jumps of $ {\bf Z}_t $ that proceed without a simultaneous jump of $ {\bf Y}_t $.
It is determined by the conditional
 Markov chain generator $ \hat{A}_{i | j }^{(\alpha)} (t) $.
This generator plays a special role in our construction.
It is special because the conditional
 Markov chain
generator $ \hat{A}_{i | j }^{(\alpha)} (t) $ is the {\it only}
generator
in \eqref{Markov_gen_2_2} that impacts prices of European vanilla options,
while prices of exotic options will in general depend on all generators that
enter \eqref{Markov_gen_2_2}. As will be
explained in more detail below, this is due to the following relations that follow
as long as $ \sum_{\beta} \hat{Q}_{\alpha  \beta}^{(i)}  =  0 $ and
$ \sum_{\beta} \tilde{Q}_{\alpha \beta}^{(m)}  = 0 $:
 \bea
 \label{Ahat}
 \sum_{\beta} A_{i \alpha | j  \beta} (t) &=&  \hat{A}_{i  j }^{(\alpha)} (t),
 \\
 \sum_{\beta} p_{i \alpha | j  \beta} (t)
 &=& \delta_{i j} + \hat{A}_{i  j }^{(\alpha)} (t) dt.  \nonumber
 \eea
Note that the fact that theoretical prices of vanilla options computed in USLV(2,2)
do not depend on specification of the other generators
$  \hat{Q} $ and $
\tilde{Q}$  in  \eqref{Markov_gen_2_2} has a few interesting implications.

First, it suggests a nice ``orthogonality'' property of model parameters determining
various generators that enter \eqref{Markov_gen_2_2}, such that parameters driving
prices of exotic options can be tuned (or picked) without impacting calibration
to vanillas. If prices of some exotic options are available in the marketplace, this
can be used to calibrate these two generators, after the model is calibrated to
available vanillas.

Second, in a scenario where no reliable pricing information is available for exotic
options, we could use this property of the model in order to specify a
measure of ``exoticness'' as, e.g., the amount the price of the given exotic derivative moves
under certain functional or parametric tweaks of the first two generators
in \eqref{Markov_gen_2_2}. Given two exotic options and
given tweaks to be performed on the generators in \eqref{Markov_gen_2_2} (such as a common
rescaling of all elements)
while pricing both options, one of the options from the pair would in general
end up being ``more exotic'' than the other.
While these issues will be addressed in a future work, here we concentrate
on the problem of calibrating the model to  European vanilla option prices.


\subsection{Conditional Markov Chain Generator}
\label{sect:Cond_MC_gen}

Clearly, prices of European vanilla options on a given underlying $ {\bf Z}_t $ for
a set of options maturing at times $ T_1, T_2, \ldots $
are only determined by marginal distributions of $ {\bf Z}_t $ at these times. An equation
driving evolution of marginal $ {\bf Z}$-distributions in the full USLV(2,2) model can be
obtained by summing over $ \beta = (\beta_1, \beta_2) $ in the forward equation \eqref{forward}.
We obtain
\beq
\label{forward_marginal}
\pi(j, t + dt ) = \sum_{\beta} \sum_{i, \alpha}
p_{i \alpha | j \beta } (t, t + dt)
\pi(i , \alpha, t) 
= \sum_{i, \alpha} \left( \delta_{i j}  +
\hat{A}_{i j}^{(\alpha)} (t)  dt \right)
\pi(i ,\alpha, t),
\eeq
where we used \eqref{Ahat} for the last equality.
This justifies the claim we made above: observed prices of European vanilla options
impose certain constraints on the conditional Markov chain generator
$ \hat{A}_{i j }^{(\alpha)} $, but not on the other generators appearing in
\eqref{Markov_gen_2_2}.  Rearranging terms in \eqref{forward_marginal}, we obtain
\beq
\label{forward_marg_2}
\frac{ d \pi(j, t)}{ dt} =  \sum_{i, \alpha}
\hat{A}_{i j}^{(\alpha)} (t)
\pi(i ,\alpha, t).
\eeq
An explicit expression for the conditional Markov chain generator
$ \hat{A}_{i j}^{(\alpha)} (t) $
can be obtained using \eqref{composition_law}:
\beq
\label{AhatG}
 \hat{A}_{i j}^{(\alpha)} = \frac{1}{dt}
 \left[
P \left[ {\bf Z}_{t +  dt} = {\bf Z}_j | {\bf Z}_{t} = {\bf Z}_i, {\bf Y}_{t} = {\bf Y}_{\alpha} \right]
- \delta_{i j}  \right] + O \left( dt^2 \right).
\eeq
From this point onwards, we reserve the notation
$ \hat{A}_{i j}^{(\alpha)} $
for a {\it calibrated} generator of USLV(2,2), while using the notation
$ A_{i j}^{(\alpha)} $ for an {\it initial guess}
for $ \hat{A}_{i j}^{(\alpha)} $. The latter is assumed
to be a valid
generator (obtained from some consistent model)
that is not necessarily accurately calibrated
to the observed market. In what follows, we will refer to the latter as a {\it prior} conditional
Markov chain generator. While a particular functional relation between the two generators
$ \hat{A}_{i j}^{(\alpha)}  $ and
$ A_{i j}^{(\alpha)}  $ will be considered in the next section, in the reminder
of this section we concentrate on specifying the second, ``prior'' generator
$ A_{i j}^{(\alpha)}  $.

As was outlined
above (see also Appendix B), we define $ A_{i j}^{(\alpha)}  $
as a combination of generators $ A_1 $ and $ A_2 $
(see \eqref{A_2D_two}), scaled by two components of $ {\bf Y}_t = \left( Y_t^{(1)}, Y_t^{(2)} \right) $:
\beq
\label{QYZ}
A_{i j}^{(\alpha)}
 = Y_{\alpha_1}^{(1)} A_1 + Y_{\alpha_2}^{(2)} A_2.
\eeq
Recalling the original definition \eqref{A_2D} of the Markov chain generator, we can
write this in a matrix form:
\bea
\label{A_2D_Y}
A_{i j}^{(\alpha )}  \, = \, \left( \begin{array}{clcrccc}
L_Y^{(0)}  & F_Y^{(0)} & 0 & 0 & \cdots & 0  & 0  \\
B_Y^{(1)} & L_Y^{(1)} & F_Y^{(1)}  &  0 & \cdots & 0 & 0 \\
0 &  B_Y^{(2)}  & L_Y^{(2)} & F_Y^{(2)}  & \cdots & 0 & 0 \\
\vdots                        \\  	
 0  & 0 & 0 &  0 & \cdots & B_Y^{( p )}  & L_Y^{( p )}   \\
\end{array} \right),
\eea
where all matrices $ B_Y^{(i)}, \, F_Y^{(i)} $ are obtained by scaling
of $ B^{(i)}, \, F^{(i)} $ by $ Y_{\alpha_1}^{(1)} $ :
\beq
B_Y^{(i)} = Y_{\alpha_1}^{(1)} B^{(i)} \, , \; \;
F_Y^{(i)} = Y_{\alpha_1}^{(1)} F^{(i)},
\eeq
while elements of $  L_Y^{(0)} $ are scaled by $ Y_{\alpha_2}^{(2)} $, except for the diagonal
elements:
\bea
\label{L_Y_elements}
\left( L_Y^{(i)} \right)_{jk}  \, = \, \left\{ \begin{array}{ll}
Y_{\alpha_2}^{(2)}  L_{jk} ^{(i)}
& \mbox{if $ k = j \pm 1 $},  \\
Y_{\alpha_2}^{(2)} \left(  L_{jj}^{(i)}  + \hat{F}_{jj}^{(i)} \right)
 - Y_{\alpha_1}^{(1)}  \hat{F}_{jj}^{(i)}
& \mbox{if $ k = j $},
\end{array}
\right.
\eea
where $ \hat{F}^{(i)} $ is defined in \eqref{A_2D_two}. Using \eqref{a_elements} in \eqref{L_Y_elements}, we obtain the explicit expression:
\bea
\label{L_Y_elements_2}
\left( L_Y^{(i)} \right)_{jk}  \, = \, \left\{ \begin{array}{ll}
Y_{\alpha_2}^{(2)}
 \frac{s_{ij}(t)}{2}
\left( \frac{\sigma_2^2}{ \Delta z_2^2} - \frac{
| \rho| \sigma_1 \sigma_2|}{ \Delta z_1 \Delta z_2} \right)
& \mbox{if $ k =  j \pm 1 $},  \\
- Y_{\alpha_2}^{(2)}
s_{ij}(t) \left( \frac{\sigma_2^2}{ \Delta z_2^2} - \frac{
| \rho| \sigma_1 \sigma_2|}{ \Delta z_1 \Delta z_2} \right)
 - Y_{\alpha_1}^{(1)}  s_{ij}(t) \frac{\sigma_1^2}{ \Delta z_1^2}
& \mbox{if $ k = j $}.
\end{array}
\right.
\eea
Clearly, conditional on values $ Y_{\alpha_1}^{(1)}, Y_{\alpha_2}^{(2)} \geq 0  $,
diagonal elements of the conditional Markov chain generator \eqref{A_2D_Y} given by
the second line of \eqref{L_Y_elements} are negative (as
long as \eqref{Delta_y_constraint} holds),
while
all off-diagonal elements are positive, and the row-wise sums of elements
in \eqref{A_2D_Y} are all zeros.
Therefore, \eqref{A_2D_Y} is a valid
conditional generator for any fixed values of $ Y_{\alpha_1}^{(1)}, Y_{\alpha_2}^{(2)} \geq 0  $.
The first component $ Y^{(1)} $ modulates transitions between
$Z^{(1)}$-states  (which may or may not be accompanied by transitions between
$Z^{(2)}$-states), while the second component $ Y^{(2)} $ modulates transitions between
$Z^{(2)}$-states  without simultaneous transitions between
$Z^{(1)}$-states.

\subsection{Fast Calibration of USLV(2,2) by 1D Forward Induction}
\label{sect:BJN}


In this section, we present a fast calibration algorithm that enables
a recalibration of the full 2D USLV(2,2) model starting from a calibrated 1D
USLV(2,0) without using a forward induction on a full-blown 2D Markov chain. It
uses a recursive procedure of ``integrating in'' the stochastic volatility process.
Our method is similar to \citet{AH}.
In its turn, a fast calibration method on a 2D Markov chain used in
\citet{AH}
is similar to an algorithm originally developed by
Britten-Jones and Neuberger (BJN)\footnote{
This approach was later popularized by \cite{Piterbarg2006}
under the name ``Markovian projection.'' Note that both BJN and Peterbarg cite the work by Dupire on the
link between stochastic and local volatility models. Dupire's
approach seems to provide a common basis for both the BJN and Markov
projection methods.}.

Recalling our previous notation where we used symbols $ \hat{A} $ and $ A $ for the
calibrated and ``prior'' conditional Markov chain generator, we assume the following relation
between them:
\beq
\label{AhatA}
\hat{A}_{i j}^{(\alpha)} (t) =
\left( 1 - \delta_{i j}  \right) q_{i j}({\bf Y}_t, t)
A_{i j}^{(\alpha)} (t) 
- \delta_{i j}  \sum_{m \neq 0}
q_{j m}({\bf Y}_t, t)
A_{j   m}^{(\alpha)} (t).
\eeq
Here $ q_{i j}({\bf Y}_t, t) \geq 0 $ are adjustment factors that will be used below to calibrate
the full-blown USLV(2,2) model.\footnote{The theoretical
interpretation of adjustment
factors $ q_{i  j}({\bf Y}_t, t) $
is that they provide ``risk-neutralizing'' drift corrections to the dynamics in the presence of
stochastic volalitity;
see a related discussion in \citet{BJN} and \citet{Rossi}.} Note
that \eqref{AhatA} defines a valid generator as long as
$ q_{i  j}({\bf Y}_t, t) \geq 0 $ and  $
A_{j  m}^{(\alpha)} (t) $ is a valid generator, as all nondiagonal elements
of $ \hat{A}_{i   j}^{(\alpha)} (t)  $ are non-negative, all diagonal elements
are negative, and all rows sum up to zero.

The purpose
of introducing the adjustment factors $ q_{i  j}({\bf Y}_t, t) $
in \eqref{AhatA} is
to provide
degrees of freedom needed for calibration
to option prices
in the (2,2) model in a way similar to the way the 1D speed factors were used above to
calibrate the (2,0) model without stochastic volatility.
As will be shown below, such calibration can be done in a
numerically efficient way by
reutilizing results of a previous calibration in a local
volatility USLV(2,0)
model.
Note that after calibration of USLV(2,2) is done via a choice of
multiplicative adjustment
factors  $ q_{i  j }({\bf Y}_t, t) $, the latter
can be combined with the 1D SFs
$ s_{ij}(t) $ that appear in the ``prior'' generator $
A_{j   m }^{(\alpha)} (t) $
 to produce 2D SFs $ S_{ij}(Y_t, t) $.

To proceed, we first plug \eqref{AhatA} into \eqref{forward_marg_2}. This yields
\beq
\label{margin_eq_22}
\frac{ d \pi(j, t)}{ dt} =  \sum_{ i \neq j, \alpha}
\left[ q_{i  j }({\bf Y}_t, t)
A_{i j}^{(\alpha)} (t)
\pi(i ,\alpha, t)  -
q_{j  i}({\bf Y}_t, t)
A_{j  i}^{(\alpha)} (t)
\pi(j ,\alpha, t) \right].
\eeq
This can be compared to the forward equation obtained in the USLV(2,0) model where we have
the following definition of the generator:
\beq
p_{i j} (t,t+dt) = \delta_{i j}  + A_{i j} dt +
O \left( dt^2  \right),
\eeq
while the forward equation has the form
\beq
\label{margin_eq_20}
\frac{ d \pi(j, t)}{ dt} =  \sum_{i \neq j}
\left[
A_{i  j} (t)
\pi(i ,t) -
A_{j   i} (t)
\pi(j, t) \right].
\eeq
Comparing \eqref{margin_eq_22} and \eqref{margin_eq_20}, we find that marginal distributions
$ \pi(j, t + dt) $ in both the USLV(2,2) and USLV(2,0) models are matched
at each node $ j = (j_1, j_2) $ provided we make
the following choice for adjustment factors $  q_{i  j}(t) $ in \eqref{AhatA}:
\beq
\label{BJN_solution}
q_{i  j}({\bf Y}_t, t)  = \frac{  A_{i j} (t)
\pi(i ,t) }{
\sum_{\alpha}
A_{i j}^{(\alpha)} (t)
\pi(i ,\alpha, t) } =
\frac{  A_{i  j} (t)
\sum_{\alpha}
\pi(i, \alpha, t) }{
\sum_{\alpha}
A_{i j}^{(\alpha)} (t)
\pi(i ,\alpha, t) }.
\eeq
We now use our key result \eqref{BJN_solution} to set up a convenient and
fast forward-induction method for the calibration of USLV(2,2)
that utilizes the results of the 1D calibration of the $Z_t$-Markov chain of the
USLV(2,0) model with a calibrated generator $ A_{i  j} (t)  $, starting with
a ``prior'' conditional generator $ A_{i j}^{(\alpha)} (t) $
of the USLV(2,2) model.

We assume that the 1D calibration of the $Z_t $-Markov chain is performed as
discussed above. We start
with the initial conditions for the 2D and 1D probability distributions, correspondingly,
 \beq
 \label{init_cond}
 \pi(i, \alpha, 0) = \delta_{i \hat{i}}
 \delta_{\alpha \hat{\alpha}}  \, , \; \;
  \pi(i, 0) = \delta_{i \hat{i}},
 \eeq
 where $ \hat{i} = ( \hat{i}_1, \hat{i}_2) $
and $ \hat{\alpha} = ( \hat{\alpha}_1, \hat{\alpha}_2 ) $ are indices
corresponding to the initial values of
 $ Z_0 $ and  $ Y_0 $ (which we assume to be
 known), respectively.
 Using \eqref{BJN_solution}, we solve for
$ q_{\hat{i}  j }({\bf Y}_0, 0) $:
 \beq
 \label{init_q}
 q_{\hat{i} j}({\bf Y}_0, 0)  = \frac{  A_{\hat{i} j} (0)
 }{
A_{ \hat{i}   j}^{(\hat{\alpha})} (0) }.
\eeq
 Note that for $ i \neq \hat{i} $, the correction factors at time $ t = 0 $ are undefined.
 However, this does not
 pose any problem as such states are unachievable at time $ t = 0 $, and therefore play no
 role in the dynamics. If desired, these parameters can be assigned
 some dummy
 values that would not have any impact on any numerical results produced with the model.

 Next we use the forward equation on interval $ [0,dt] $ to compute the
joint probability  $ \pi(j,\beta, dt) $, which is then used to compute the
adjustments for all nodes at time $ t = dt $, and so on.
As a result, we have a full 2D USLV(2,2) Markov chain
calibrated to the set of option quotes using a fast and effective
algorithm.

\subsection{Generator of Idiosyncratic Dynamics of $ {\bf Y}_t $}
\label{sect:Discrete_Y_dynamics}

In this section, we provide some examples of specification of the first generator,
$ \delta_{ij} \hat{Q}_{\alpha  \beta}^{(i)} $,  in \eqref{Markov_gen_2_2}.
We recall that by construction of the model, the choice of this generator in USLV(2,2)
has no impact on the quality of calibration of the model to prices of European vanilla
options in a calibration set, while in general it does impact
prices of exotic options produced by the model.

As was mentioned above, for practical applications we typically have in mind a
low (e.g., 3 to 11)
number of possible states per dimension of the stochastic volatility factor. This has two
implications.

First, we prefer to view continuous-space models as a convenient and compact way to
parametrize
the generator $ \delta_{ij} \hat{Q}_{\alpha  \beta}^{(i)}  $
in \eqref{Markov_gen_2_2}. This generator corresponds to idiosyncratic moves of $ {\bf Y}_t $
without simultaneous moves of $ {\bf Z}_t $.  Such parametrization is clearly preferred to directly specifying $ \sim 2 (2q +1)^2 $ free parameters defining a
discrete-state generator
$  \delta_{ij}  \hat{Q}_{\alpha  \beta}^{(i)}  $.

Second, as long as the number of volatility states is low,
the generator matrix
for $ {\bf Y}_t $ should not necessarily be sparse. This remark is important as
nonsparse matrices arise while discretizing jump-diffusion processes.

For simplicity, in this paper we restrict ourselves to a particular bivariate
mean-reverting diffusion process as a continuous-space model that produces
generator $ \hat{Q}_{\alpha  \beta}^{(i)} $
after a proper discretization.\footnote{If any other model
of stochastic volatility is chosen instead of the one presented below, the only
change needed
in the present framework would be to construct different transition matrices
(or generators) for the $ Y$-states, while the computational
part of calibration and pricing would stay the same.}
More specifically, we consider a bivariate Ornstein-Uhlenbeck (OU) process for
the logarithmic variables $ y_t^{(i)} = \log Y_t^{(i)} $:
\bea
\label{cond_diffusion_2D}
&& d y_t^{(1)} =
  k_1 \left( \eta_1 - y_t^{(1)} \right) dt + \nu_1 d W_t^{(1)}, \nonumber \\
&& d y_t^{(2)} =
  k_2 \left( \eta_2 - y_t^{(2)} \right) dt + \nu_2 d W_t^{(2)},
\eea
where the two Brownian motions $ W_t^{(1)} $ and $ W_t^{(2)} $ are correlated with correlation
$ \rho_y $.

The continuous-space Markov generator corresponding to
\eqref{cond_diffusion_2D} reads
\beq
\label{Markov_generator_Y_2D}
\mathcal{L}_y V( {\bf y}) = b_1(y) \frac{ \partial V(\bf y)}{\partial y_1} +
b_2 (y) \frac{ \partial V(\bf y)}{\partial y_2} + \frac{1}{2}
\nu_1^2 \frac{\partial^2 V(\bf y)}{
\partial y_1^2} + \frac{1}{2} \nu_1^2 \frac{\partial^2 V(\bf y)}{
\partial y_1^2}  + \rho_y \nu_1 \nu_2 \frac{ \partial^2 V(\bf y)}{\partial y_1 \partial y_2},
\eeq
where
\beq
\label{drift_coeff_2D}
b_i (y)  =  k_i \left( \eta_i  - y_t^{(i)} \right) \, , \; \; i = 1,2.
\eeq
Note that parameters of generator $ \mathcal{L}_y $ can be made dependent
on the value of $ {\bf Z}_t $ if desired.

The $Y$-generator can be discretized in a similar way to
a procedure used above
in Sect.~\ref{sect:USLV_N2}. We use central differences for the
second derivatives and a noncentral difference for the mixed derivative in
\eqref{Markov_generator_Y_2D}. In addition, we use the upwind-difference discretization
for the first derivatives in \eqref{Markov_generator_Y_2D}. Regrouping different
terms in the discretized generator much as it was done above in
Sect.~\ref{sect:USLV_N2}, the resulting discrete Markov chain
generator for $ {\bf Y}_t $
can be cast in the form of another QBD process, in an analogous way to our
construction of the QBD process for $ {\bf Z}_t $. The resulting generator takes the
familiar QBD form\ednote{Provide explicit formulae here for matrices $ L, F, B $.}
(compare, e.g., with \eqref{A_2D_Y}):
\bea
\label{Q_2D_Y}
\hat{Q}^{(i)}  \, = \, \left( \begin{array}{clcrccc}
L^{(0)}  & F^{(0)} & 0 & 0 & \cdots & 0  & 0  \\
B^{(1)} & L^{(1)} & F^{(1)}  &  0 & \cdots & 0 & 0 \\
0 &  B^{(2)}  & L^{(2)} & F^{(2)}  & \cdots & 0 & 0 \\
\vdots                        \\  	
 0  & 0 & 0 &  0 & \cdots & B^{( 2q )}  & L^{( 2q )}   \\
\end{array} \right),
\eea
where all matrices $ B^{(i)}, L^{(i)},\, F^{(i)} $ have dimension
$ (2q+1)\times (2q+1)$, i.e., the dimension of our one-dimensional grids.

We would like to conclude this section with a few remarks.
Using a diffusive prototype is certainly not the only way of constructing
the idiosyncratic generator $ \hat{Q}^{(i)} $ in \eqref{Markov_gen_2_2}.
Alternatively, we could consider more general jump-diffusion or L{\'e}vy processes
as continuous-space prototypes of the generator. For such more general specifications,
generator $ \hat{Q}^{(i)} (t) $ would not in general
be sparse. Note, however,
that a potential nonsparsity of the generator in the latter
case would not be a major concern if we have a small number of states
for $ {\bf Y}_t $.


Furthermore, as was mentioned above, a BJN-like fast 1D calibration
procedure for AR-USLV(2,2) presented below in Sect.~\ref{sect:BJN} can also be applied
when instead of {\it spot} variance factors $ {\bf Y}_t $, we use {\it integrated} variance
drivers/subordinators $ {\bf T}_t $. The only difference from the case of spot variance
factors just presented would be that generators for subordinators should have all zeros
below the diagonals (as a subordinator should be
a non-decreasing function of time). However, both the calibration and pricing algorithms
would stay the same.

\subsection{Generator of Joint Jump Dynamics}
\label{sect:Joint_jump}

Here we consider the second term $ ( 1 - \delta_{ij}) \tilde{Q}_{\alpha \beta}^{(j-i)}
\hat{A}_{ij}^{(\alpha)} $ in \eqref{Markov_gen_2_2}. We recall that
this expression defines the generator of joint jumps whose
matrix elements give
 intensities of simultaneous jumps of $ {\bf Z}_t $ and $ {\bf Y}_t $.

The motivation for introducing joint jumps of $ {\bf Z}_t $ and $ {\bf Y}_t $
is to incorporate the asset-volatility codependence\footnote{
Such as  the well-known
leverage effect (a negative spot-volatility correlation) for equity markets.} in our
framework. We note that in a discrete-time setting, both
\citet{BJN} and \citet{Rossi} introduce different
transition matrices
for the $ Y $-states for different transitions between the $ Z$-states in an {\it ad hoc}
way.  Our approach, which starts with a continuous time dynamics and the decomposition
\eqref{Markov_gen_2_2} of the Markov generator,
is hopefully a bit more systematic and easier to relate to one's intuition.

As the conditional Markov chain generator $ \hat{A}_{ij}^{(\alpha)} $
is fixed by the above
procedure of calibration to
vanilla options, the joint jump generator is specified by defining
the pseudo-generator $ \tilde{Q}_{\alpha \beta}^{(m)} (t) $ (see
\eqref{cond_gen}).
Much as we did in our approach above for the idiosyncratic $Y$-generator, here we settle for
a very simple and parsimonious choice. Alternative and
more complicated specifications of the pseudo-generator $ \tilde{Q}_{\alpha \beta}^{(m)} (t) $
 could clearly be considered
instead without significantly affecting complexity or performance of the model.

Specifically, we assume that
conditionally on $ \left( \Delta Z_t^{(1)}, \Delta  Z_t^{(2)} \right) $, jumps
$ \Delta  Y_t^{(1)} $
and $ \Delta Y_t^{(2)} $ are independent. We further specify that
jump  $  \Delta  Y_t^{(1)} $ depends only on $ \Delta Z_t^{(1)} $
rather than on the pair $ \left( \Delta Z_t^{(1)}, \Delta  Z_t^{(2)} \right) $, and similarly
$  \Delta  Y_t^{(2)} $ depends only on $ \Delta Z_t^{(2)} $.
To control the amount of the asset-volatility codependence in our model, we introduce
two parameters, $ \gamma_1 $ and $ \gamma_2 $, that determine the codependence for
the joint moves $ \left( \Delta Z_t^{(1)}, \Delta Y_t^{(1)} \right) $ and
$ \left(\Delta  Z_t^{(2)}, \Delta  Y_t^{(2)}
\right) $, respectively, such that $ \gamma_i > 0 $ and $ \gamma_i < 0 $ correspond to
positive and negative codependences, respectively.
 We then define the pseudo-generator
$ \tilde{Q}_{\alpha \beta}^{(m)} (t) $ as follows:\footnote{This parameterization was proposed by Leonid Malyshkin.}
\bea
\label{Atilde}
\tilde{Q}_{\alpha \beta}^{(m)} (t)
&=& \left[ \delta_{\alpha_1 \beta_1}
+ \left( \gamma_1 m_1 \right)^{+} A_{\alpha_1 \beta_1}^{(1,+)}
+ \left( \gamma_1 m_1 \right)^{-} A_{\alpha_1 \beta_1}^{(1,-)} \right]   \\
& \times &
\left[ \delta_{\alpha_2 \beta_2}
+ \left( \gamma_2 m_2 \right)^{+} A_{\alpha_2 \beta_2}^{(2,+)}
+ \left( \gamma_2 m_2 \right)^{-} A_{\alpha_2 \beta_2}^{(2,-)} \right]
- \delta_{\alpha_1 \beta_1} \delta_{\alpha_2 \beta_2}  \, , \; \;
(m_1, m_2) \neq (0, 0),
\nonumber
\eea
where for any real number $ x $ we defined $ x^{+} = \max(0,x) $ and $ x^{-} =
\max(0, -x) $. Matrices $ A^{(+)} $ and $ A^{(+)} $ in \eqref{Atilde} stand for
some upper- and lower-triangular generator matrices, respectively.
For example, a nonsparse specification of generator \eqref{Atilde}
could be provided using two parameters,
$ q_{1}, q_{2} \leq 1 $, that determine the speed of decay of the generator away from
the diagonal:
\bea
\label{Apm}
A_{\alpha \beta}^{(i,+)} &=& \theta( \beta - \alpha) q_i^{\beta - \alpha - 1}
- \delta_{\alpha \beta} \sum_{\beta > \alpha} q_i^{\beta - \alpha - 1}
\;, \; \; i = 1,2,  \nonumber \\
A_{\alpha \beta}^{(i,-)} &=& \theta( \alpha - \beta) q_i^{\alpha - \beta - 1}
- \delta_{\alpha \beta} \sum_{\beta < \alpha } q_i^{\alpha - \beta -  1}
 \;, \; \; i = 1,2,
\eea
where $ \theta(x) $ stands for a Heavyside step-function. Alternatively, if we want
to keep the generator sparse, we could consider bi-diagonal specifications for
matrices $ A^{(i,\pm)} $.

\section{ITC-USLV(2,2): Implied Time Change Process}
\label{sect:USLV22_with_ITC}

 The algorithm of forward induction-based calibration of the USLV model just presented
is simple
 and intuitive; however, it is not ideal from a practical viewpoint, as it assumes that
the time steps are small enough to justify the use of a trinomial (birth-and-death)
approximation for the diffusion process in $ {\bf Z}_t $. In practice, this
means we should take daily (possibly hourly) steps, which
may slow down calibration and pricing. If we want to be able to have a lattice with
larger steps (e.g., monthly), or work with irregular large steps, we need a
different method.

Several alternatives of different complexity can be considered at this point.
One approach would be to generalize \eqref{AhatA} to the case of larger time
steps $ \Delta t $ by viewing  the left- and right-hand sides
of \eqref{AhatA} as leading terms in expansions of
finite-time matrix exponentials of the conditional (on a realization of $ {\bf Y}_t $)
generator of the QBD Markov chain for the calibrated and ``prior'' model,
respectively. While it can be shown that the recursive forward
calibration can be carried over in such framework as in the one-step BJN method
described above, practical uses of such an approach may be constrained by our ability to
compute conditional multi-step transition probabilities for $ {\bf Z}$-states
in a numerically efficient way.
We expect that splitting methods can be efficiently used to this end, but we leave
research in this direction for a future work, and instead concentrate on alternative
approaches.  The latter are based on stochastic time change techniques, which is what we
present next.


\subsection{Modeling Stochastic Time Changes}
\label{sect:stoch_time_changes}

Let $ \xi_t $ be a (random) matrix-valued value of the QBD process with generator \eqref{A_2D} at time $ t $.  Consider a right-continuous nondecreasing process
$ T_t $ with $ \tau_0 = 0 $ with independent and homogeneous increments.\footnote{That is,
 $ \tau_{t+s} - \tau_{t} $ is independent of the filtration $ \mathcal{F}_t $ and has the same
distribution as $ \tau_s $.} In the present context,
such process $ T_t $ is called a {\it Bochner subordinator};
see, e.g., \citet{Feller}.

Now consider a new process $ \eta_t \equiv \xi_{T_t} $ given by our QBD process \eqref{A_2D} evaluated at a random (business) time  $ T_t $ instead
of the calendar time $ t $. This produces a QBD Markov chain {\it subordinated} to the Bochner
subordinator $ T_t $.

Note that the idea of a subordinated Markov chain has already been used above in
Sect.~\ref{sect:randomization} (see also a discussion on this point in \citet{GM}) as a
computational tool for evaluation of matrix exponentials of generator \eqref{A_2D}.
A more general subordinator is given by a nondecreasing L{\'e}vy process; see, e.g., \citet{CGMY}
and references therein. It can be written as
\beq
\label{time_change_gen}
T_t = \int_{0}^{t} Y_s ds + T_t^{(jump)},
\eeq
where $ Y_t $ is a nonnegative process called the {\it activity rate}, and $ T_t^{(jump)} $
stands for a jump component of the time change.

Many specifications of a time change process can be described by the general formula
\eqref{time_change_gen}. For example, subordination by a Poisson process was used above in
Sect.~\ref{sect:randomization}, which corresponds to the Bochner subordinator being a pure jump
process with a finite jump activity. Another simple choice for a pure jump time change would be
a gamma process (incidentally, this process has a particularly simple Laplace transform).
Alternatively, one can consider purely diffusive
time changes where $ T_t^{(jump)} $ vanishes, with the activity rate $ Y_t $ specified, e.g.,
by a CIR process or a positive OU process.

Let us assume that the stochastic time change is independent of
the QBD Markov chain, and that its Laplace transform
\beq
\label{Laplace}
\mathcal{L}_{T_t} (u) = \mathbb{E} \left[ e^{ - u T_t} \right]
\eeq
is known in closed form, or can be computed numerically at a low cost.
Consider first a one-factor stochastic time change for a single Markov chain with generator
$ A $. Recall that finite-time transition probabilities can be computed using the randomization
method as in \eqref{Taylor_P}, which we write here as
\beq
\label{Taylor_P_2}
P(t) = P_0' e^{- \lambda t } \sum_{n=0}^{\infty} \frac{
\left( \lambda t \right)^n {\bf P}^n}{n!}
 =  \sum_{n=0}^{\infty} \frac{(-1)^n}{n!}  P_0' {\bf P}^n \left( \lambda \frac{d}{d \lambda} \right)^n
e^{- \lambda t }.
\eeq
Substituting here $ t \rightarrow T_t $, taking
the expectation with respect to future scenarios of the time change process $ T_t $, and
interchanging the summation and expectations in \eqref{Taylor_P_2}, we obtain
\beq
\label{Taylor_P_3}
\mathbb{E} \left[ P(T_t)\right] =
\sum_{n=0}^{\infty} \frac{(-1)^n}{n!} P_0' {\bf P}^n \left( \lambda \frac{d}{d \lambda} \right)^n
\mathcal{L}_{T_t} (\lambda).
\eeq
Note that in practice, randomization methods truncate an infinite series in
\eqref{Taylor_P_2} at some finite $ n_{max} $ determined by a needed accuracy level
$ \varepsilon $.
If we first truncate the series for fixed $ t $ given $ \varepsilon $,
and {\it then} do a stochastic time change, this might lead to a substantial loss of
accuracy. One possible way to achieve a fixed-$ \varepsilon $ calculation for a given $ t $
in the model with stochastic time $ T_t $ is as follows. We first find a maximum
value of $ \tau_{max} = \max_{\tau} T_t $ that can be reached at some confidence level, and
then truncate the sum at some value $ n_{max}' $ that provides needed accuracy for
\eqref{Taylor_P_2} where $t$ is replaced by $ \tau_{max} $.

This means that if derivatives of the Laplace transform of the time change are easy to
compute, and the number of terms we need to keep in \eqref{Taylor_P_3} for given tolerance
is reasonably small, then the problem of parametric calibration of
parameters of the $Y$-process
in the USLV(1,1) (or USLV(2,1), see below) models can be solved, in the
zero-correlation limit, in one step, with no need for a forward induction algorithm.

Note that while the assumption of zero correlation might be restrictive, the above approach can
in fact be generalized to the case of nonzero correlation, at the price of
introducing a complex-valued measure (see \citet{CW_2004}). This could be used to calibrate parameters
of the $ Y $ process for a given set of option quotes while keeping the 1D SFs fixed or flat.\footnote{Alternatively, we could use the zero-correlation limit as a ``quick and dirty'' way to
estimate the parameters of the full model with nonzero correlation, except of course those parameters
that are critically dependent on the level of correlation.} Further
improvements of calibration quality (if desired)
could then be achieved using a forward-induction-based calibration method described in
Sect.~\ref{sect:BJN}.


For the most interesting {\it two-factor} time change specification (i.e., for USLV(2,2)), the situation is more
tricky because of correlations between different drivers, as well as because of noncommutativity
of generators $ A_1 $ and $ A_{2} $. However, as will be shown in the next section, it turns out that a tractable framework can be
obtained with a proper construction of a two-factor time change.


\subsection{Time Change with a Hierarchical Bivariate Subordinator}

It might be tempting to try to extend our framework by incorporating stochastic time changes
with a nonvanishing jump component $ T_t^{(jump)} $ in Eq.(\ref{time_change_gen}),
with a two-factor time change $ {\bf T}_t = ( T_t^{(1)}, T_t^{(2)} ) $.
If $  T_t^{(1)},  T_{t}^{(2)} $ include jumps, this leads to jumps in the
underlyings $ {\bf Z}_t = \left( Z_t^{(2)}, Z_t^{(2)} \right) $. If both the nonvanishing
activity rate
$ {\bf Y}_t $ and a jump component   $ \hat{T}_t^{(jump)} $ are retained
in Eq.(\ref{time_change_gen}), this results (in our setting) in a four-factor stochastic-local
volatility dynamics with jumps in both the underlyings and volatility.

Now we introduce such a two-factor stochastic time change with a bivariate subordinator
and show how to evaluate finite-time transition probabilities in
a resulting subordinated Markov chain using a generalization
of the randomization method presented in Sect.~\ref{sect:randomization}.

Recall from Sect.~\ref{sect:stoch_vol} that the Markov chain generator $ A $
in our problem
has the decomposition \eqref{A_2D_two}, where both $ A_{1} $ and $A_{2} $
are valid
generators that can be subject to individual stochastic time changes.
Consider a bivariate
subordinator of the following form
\bea
\label{tau_2D}
T_t =
\left( \begin{array}{cc}
 T_t^{(1)} \\
T_t^{(2)}\\
\end{array}
\right) =
\left( \begin{array}{c}
 \theta_t T_t  \\
 T_t   \\
\end{array}
\right).
\eea
Here $ \theta_t > 0 $ is a stochastic process with
$ \mathbb{E} [ \theta_t ] = 1 $ that
will be
specified in more detail below. We assume that $ \theta_t $ is
independent of $ T_t $.
As $ \theta_t $ acts as a time dilation factor on top of the random time
$ T_t $, we
will refer to $ \theta_t $ as a {\it time-dilaton} process.
Note that correlation between $ T_t^{(1)} $ and $ T_t^{(2)} $ is now driven by
the variance of $ \theta_t $:
\beq
\label{corr_tau}
\rho_{T_t^{(1)}, T_t^{(2)}} = \sqrt{ \frac{ Var \left( T_t \right)}{
  Var \left( T_t \right) + Var \left( \theta_{t} \right) \left(
Var \left(
T_t \right) + \left( \mathbb{E}\left[ T_t \right] \right)^2 \right)}}
\eeq
so that we can fit any nonnegative correlation between $ T_t^{(1)} $ and
$ T_t^{(2)} $
by a proper choice of $ Var (\theta_t) $.

We assume that $ (T_t, \theta_t) $ is a 2D Markov process with independent and
time-homogeneous
increments. Furthermore, we assume that both $ T_t $ and $ \theta_t $ are
non-decreasing exponential-L{\'e}vy processes. As a product $ \theta_t T_t $ of two
(non-decreasing) exponential-L{\'e}vy processes $ T_t $ and $ \theta_t $ is another
(non-decreasing)  exponential-L{\'e}vy process, \eqref{tau_2D} defines a valid subordinator
that
can be used to
time-change our QBD process with generator \eqref{A_2D_two}.

The interpretation of the bivariate subordinator \eqref{tau_2D} is as
follows. The second
component
$ T_t^{(2)} = T_t $ provides a common time change that modulates all transitions
on the chain; i.e., it acts on both generators $ A_1 $ and $ A_2 $. The first component
$ T_t^{(1)} = \theta_t T_t $ can be thought of as a {\it hierarchical} time
change. In this hierarchical scheme,
we first apply a common time change $ T_t $ , and then time-change it
again
using a linear time change function
$ T_t^{(1)}(T_t) = \theta_t T_t $. This time change will be applied below to
generator $
A_1 $ alone.\footnote{Note that
the order of
the time
changes indicated above is very important: if we reversed it, this would be
equivalent
to allowing future events of transitions driven by $ A_1 $ to impact the
dynamics of the whole system at the
present time.
For a recent application of
such hierarchical time changes, see, e.g., \citet{Puzanova_2011}.} Note that if
both $ T_t $ and $ \theta_t $ are non-decreasing
exponential L{\'e}vy processes, this implies that the stochastic clock runs faster
for transitions involving changes of both $ Z^{(1)} $ and $ Z^{(2)} $ than for transitions
that only involve changes of $ Z^{(2)}$.
Also note that while $ \theta_t $ is a stochastic
{\it process}, in the context of calculation of finite-time transition
probabilities on
a fixed
interval $ t \in [0, t] $ (where $ t $ is some
``interesting'' time, e.g., a coupon date) what matters is only a terminal value
$ \theta_t $ (see below
in \eqref{finite_time_P_tc}). Therefore, for such calculation we can treat
the terminal
value
$ \theta_t $ as a random variable,\footnote{This is due to the Markov property of
the $ \theta_t $-dynamics which was assumed above: For a continuous-time Markov process,
a time line can be chosen in an arbitrary way, while the corresponding finite-time
 transition probabilities would be related by the Chapman-Kolmogorov equations; see, e.g., \citet{Feller}.}
 which certainly simplifies an approach presented below.

The finite-time transition probability for a time-changed QBD Markov chain
can now be
computed by conditioning on the realization of $ \theta_t $:
\bea
\label{finite_time_P_tc}
P (t) &=& \mathbb{E} \left[ P_0' e^{ T_t^{(1)} A_{1} + T_t^{(2)} A_2 } \right]
= \mathbb{E} \left[ P_0' e^{ \theta_t T_t A_{1} + T_t A_2 } \right]
= \mathbb{E} \left[  \left.
\mathbb{E} \left[ P_0' e^{ T_t \left( \theta_t  A_{1} + A_2 \right)} \right]
\right|
\theta_t \right]
\nonumber  \\
&\equiv&
\mathbb{E} \left[  \left.
\mathbb{E} \left[ P_0' e^{ T_t A_{\theta}}  \right| \theta_t \right]
\right].
\eea
Here the outside expectation corresponds to averaging with respect to the
randomness due to $ \theta_t $,
while the inner expectation averages over the randomness due to $ T_t $
for a fixed value of $ \theta_t $. In the last equation,
we have defined $ A_{\theta} = \theta_t  A_{1} + A_{2} $ for any fixed
$ \theta_t = \theta $.

Now consider the inner expectation in \eqref{finite_time_P_tc}. For any fixed
$ \theta_t = \theta $,
we proceed as follows. First we specify a nonnegative parameter
$ \Lambda_{\theta}
\geq \max_{n} \left| \left( A_{\theta} \right)_{nn} \right| $. Next, we use
the idea
of the randomization
method of Sect.~\ref{sect:randomization} to define a DTMC with transition
matrix
\beq
\label{P_mat_alpha}
 {\bf P}_{\theta} = {\bf I} + \frac{{\bf A}_{\theta}}{\Lambda_{\theta}} \;
\Rightarrow
{\bf A}_{\theta} =
\Lambda_{\theta} \left( {\bf P}_{\theta} - {\bf I} \right).
 \eeq
 Substitute $ A $ as given by \eqref{P_mat_alpha} into the solution of the
forward
 equation, which we write here as $ P_{\theta}(t) $ to emphasize that the
whole calculation
 is done for a fixed $ \theta_t = \theta $:
 \[
 P_{\theta}(t) = \mathbb{E} \left[ \left. P_0' e^{T_t A_{\theta}} \right|
\theta_t
\right] =
 \mathbb{E} \left[ \left.  P_0' e^{ T_t  \Lambda_{\theta} \left(
{\bf P}_{\theta} -
{\bf I} \right)}
 \right| \theta_t  \right] =
 \mathbb{E} \left[  \left.  e^{- \Lambda_{\theta} T_t }
 P_0' e^{  T_t \Lambda_{\theta} {\bf P}_{\theta} } \right| \theta_t  \right].
 \]
 Using a Taylor series expansion for the matrix exponential in this expression
 and interchanging the summation and expectation, we obtain
 \beq
 \label{Taylor_P_theta}
 P_{\theta}(t) = \sum_{n=0}^{\infty} \frac{\left( P_0' {\bf P}_{\theta}^n
\right)}{
n!}  \mathbb{E} \left[ \left.
    \left( \Lambda_{\theta} T_t \right)^n e^{- \Lambda_{\theta} T_t }
 \right| \theta_t \right] =
 \sum_{n=0}^{\infty} \frac{(-1)^{n}}{n!} \left( P_0' {\bf P}_{\theta}^n
\right)
 \left. \left( \lambda \frac{d}{d \lambda} \right)^n
\mathcal{L}_{T_t}(\lambda)  \right|_{\lambda = \Lambda_{\theta}},
 \eeq
where $ \mathcal{L}_{T_t} (u) $ stands for a Laplace transform
\eqref{Laplace} of
the time change $ T_t $.

\eqref{Taylor_P_theta} is a ``semi-analytical'' expression for a finite-time
transition
probability in
our Markov chain {\it after} the first time change driven by $ T_t $, but
{\it before}
the second
time change driven by $ \theta_t $. The product $  P_0' {\bf P}_{\theta}^n $
can be
efficiently implemented via a recursive vector-matrix multiplication as in
\eqref{recursive_PP}.
The derivatives $ \left( \lambda \frac{d}{d \lambda} \right)^n
\mathcal{L}_{T_t}(\lambda)$
can be easily computed if the Laplace transform
$ \mathcal{L}_{T_t} ( \lambda) $ is known in closed form (or can be computed numerically
at a low cost). Note that this
part of the calculation is
independent of
any pricing data---the latter only impacts the matrix $ {\bf P}_{\theta} $ through a calibrated
set of
1D SFs $ s_{i} $.

As an example, consider an exponential-gamma subordinator specification for
$ T_t^{(1)} = T_t $ with $ T_t = \exp(X_t ) $, where $ X_t $ is a  Gamma process.
Recall that an univariate (homogeneous) Gamma process
$ X_t \geq 0 $ with $ X_0 = 0 $ and parameters $ a,b \in \mathbb{R}^{+} $
is a process with independent increments such that
$ X_t $  is Gamma-distributed  $ \Gamma (a t, b) $ with
the following probability density function (pdf):
\beq
\label{pdfGamma}
f_{ a t, b} (x) = \frac{b^{ a t}}{ \Gamma( a t)}
x^{ a t - 1} e^{ - bx} 1_{ \mathbb{R}^{+}}(x),
\eeq
with
\beq
\label{meanVarGamma}
\mathbb{E} \left[ X_t \right ] = \frac{ a t}{ b} \, , \; \;
Var \left[ X_t \right ] = \frac{ a t}{ b^2}.
\eeq
Here the parameter $ a $ is called the shape parameter, and $ b $ is the rate parameter.
In what follows, we set $ a = b = 1/\nu $. The Laplace transform
of \eqref{pdfGamma} reads
\beq
\label{Laplace_gamma}
\mathcal{L}_{X_t}(u) = \left( 1 + \nu u \right)^{- \frac{t}{\nu}}.
\eeq
Given this expression, we can approximately compute the Laplace transform of $ T_t = \exp(X_t) $.
All rescaled derivatives $ \left( \lambda \frac{d}{d \lambda} \right)^n \mathcal{L}_{T_t}(\lambda) $
would then have to be computed from that latter Laplace
transform\ednote{Provide some details here.}.

After the condtional time-$t$ probabilities are computed, the unconditional
probabilities are obtained by averaging over a marginal probability density
$ p_t(\theta) $ of $ \theta_t $:
\beq
\label{uncond_P_tc}
P(t) = \mathbb{E} \left[ P_{\theta}(t) \right] = \int d \theta p_t(\theta)
P_{\theta}(t).
\eeq
In practice, this integral should be computed by discretization
of the range of $ \theta_t $ onto a finite grid $
[\theta_0, \theta_1, \ldots, \theta_{q-1} ] $.

Note that we can proceed in two
different ways with a computation involved in \eqref{uncond_P_tc}. The
first way would be to specify a process for $ \theta_t $, discretize it, and
then compute a discrete approximation to \eqref{uncond_P_tc}. We could tune
parameters of this process to fit a given set of option quotes. However, a particular
parametric model of $ \theta_t $ might be too restrictive for such task,
especially if the number of option prices to fit grows larger. For this reason,
in the next section we present a more flexible nonparametric approach that is
able to fit any arbitrage-free set of option quotes.


\subsection{Implied Time-Dilaton Process}

For a particular parametric specification of a (discretized) time dilaton
process $ \theta_t $, \eqref{uncond_P_tc} produces some
transitions probabilities for the $ Z$-states in the
time-changed QBD Markov chain.
In general, these transition probabilities would be different from marginal
probabilities in
the local volatility USLV(2,0) calibrated to observed option prices. This
means that a nearly perfect calibration to options achieved in the USLV(2,0)
model would in general be lost once we add a stochastic time change to our model.

However, we can rematch the prices of options in our calibration set after a
time change
if we treat the distribution $ p_t(\theta) $ of realizations of different
values of $ \theta_t $
as a distribution {\it implied} by option prices (given the specification
of a process
for $ T_t $). Furthermore, for a multi-period setting specified by
a particular time grid $ t_0, t_1, \ldots $ (where $ t_i $ can be, e.g.,
swaption maturities in the calibration set), we can construct an
implied {\it process} for $ \theta_t $
if we impose a Markovian structure on it. In this section, we show
how such process can be constructed using a Minimum Cross Entropy (MCE)
method (see, e.g., \citet{CT}). Our construction is similar to \citet{IMFM} where analogous
ideas were used in a different context.

Let $ F_{ijk} $ be a payoff function  of the option with
maturity $ t_i $ and strike
$ K_j $ (with $ j = 1, \ldots, K $)
in a scenario where the terminal value of the underlying at the option
maturity is given by values $ \left( Z_{k_1}^{(1)}, Z_{k_2}^{(2)} \right) $, and
let $ C_{ij} $ be the corresponding market prices of options in our calibration set.
Using \eqref{uncond_P_tc}, we obtain a set of
constraints\footnote{Note that we use the continuous notation here
for simplicity of presentation only. For implementation, all stochastic processes
are discretized within our approach.}
\beq
\label{constraint1}
\int d \theta p_i(\theta) G_{ij}(\theta) = C_{ij} \, , \; \;
G_{ij}( \theta) \equiv  \sum_{k} F_{ijk} \left[ P_{\theta} \right]_{\hat{k}, k}(t_i)
\, , \; \; j = 0, \ldots, K,
\eeq
where $ \hat{k} = (\hat{k}_1, \hat{k}_2) $ is an index corresponding to the initial
value $ {\bf Z}_0 $. Note that \eqref{constraint1} with $ j = 0 $ corresponds to
the constraint  $ \mathbb{E}[\theta_t] = 1 $ implied above in
\eqref{corr_tau}, which is here enforced as an additional artificial option quote
with $ j = 0 $, $ F_{i0k} = \theta_{t} $ and $ C_{i0} = 1 $.

We can now find a probability density $ p_i(\theta) $
that satisfies these
constraints using the MCE approach. With this method, given a {\it reference} ({\it ``prior''})
model $  q_i(\theta)$
(given, e.g., by another exponential-gamma process), we minimize the
Kullback-Leibler (KL) distance between the two distributions $ p_i(\theta)$ and
$ q_i(\theta) $ (see \citet{CT}):
\beq
\label{KL}
D \left[ p_i(\theta) || q_i (\theta) \right] = \int d \theta p_i(\theta) \log \frac{ p_i(\theta)}{q_i(\theta)}
\eeq
subject to constraints of \eqref{constraint1}. This produces a least biased (relatively to the
reference measure $ q_i(\theta) $) distribution $ p_i (\theta) $ that satisfies the constraints of
\eqref{constraint1}.

For the first node on the time grid, minimization of \eqref{KL} with
constraints \eqref{constraint1} is done using the method
of Lagrange multipliers. Using \eqref{KL} with $ i = 1 $, the
corresponding Lagrange function is
\beq
\label{Lagrange1}
L = \int d \theta p_1(\theta) \log \frac{ p_1(\theta)}{q_1(\theta)} - \sum_j \xi_j^{(1)}
\left( \int d \theta p_1(\theta) G_{1j}(\theta) - C_{1j} \right),
\eeq
 where $ \xi_j^{(1)} $ are Lagrange multipliers.
 Minimizing this expression with respect to $ p_1(\theta) $, we obtain
\beq
\label{firstMat}
p_1(\theta) = \frac{1}{Z_1} q_1 (\theta)
e^{ \sum_j \xi_j^{(1)} G_{1j}(\theta) } \,, \; \; Z_1 = \int d \theta q_1 (\theta)
e^{ \sum_j \xi_j^{(1)} G_{1j}(\theta) }.
\eeq
The Lagrange multipliers
can now be computed by plugging \eqref{firstMat} back into \eqref{Lagrange1},
and maximizing the resulting expression as a function of $ \{ \xi_j^{(1)} \} $.
This amounts to a convex optimization problem
in dimension equal to the number of option quotes. (For more details on the MCE method in both the one- multi-period settings, see, e.g., \citet{IMFM} and references therein.)


For the second maturity, instead of minimizing the {\it unconditional} KL distance
\eqref{KL}, we minimize a {\it conditional} KL distance for the next interval. This is
done as follows.
Using the Markov property
we can write the pricing constraints as
\beq
\label{constraint2}
C_{2,j} = \int_{0}^{\infty} d \theta_2 \, p_2(\theta_2) G_{2j}(\theta_2) =
\int_{0}^{\infty} d \theta_2 \, G_{2j}(\theta_2)
\int d \theta_1 \, p_1(\theta_1) \, p(\theta_2|\theta_1).
\eeq
Assuming that the density $ p_1(\theta_1) $ is fixed at the
previous step, the conditional
transition density $ p(\theta_2 | \theta_1) $ can be
found by minimization of the expected
conditional KL cross entropy\footnote{The conditional KL cross entropy
is a measure of the difference between two conditional transition probabilities, averaged
over the position of the initial point; see, e.g., \citet{CT}.}
\beq
\label{conditionalCross}
H \left[ p (\theta_2|\theta_1) || q (\theta_2|\theta_1)
\right] = \int d \theta_1 \, p_1(\theta_1)
\int d \theta_2 p( \theta_2| \theta_1)
\log \frac{ p(\theta_2|\theta_1)}{ q(\theta_2|\theta_1)}
\eeq
subject to pricing constraints \eqref{constraint2}.
Here $ q(\theta_2|\theta_1) $ is a prior transition probability. As
the time change $ T_t $
should be nondecreasing, it should satisfy the condition
$ q(\theta_2|\theta_1) = 0 $ for
$ \theta_2 < \theta_1 $. Again, a natural choice for the prior transition
density could be the
transition density of an exponential-gamma process.

The corresponding Lagrange function for the second interval is
\bea
\label{Lagrange2}
L &=& \int d \theta_1 \, p_1(\theta_1)
\int d \theta_2 p( \theta_2| \theta_1)
\log \frac{ p(\theta_2|\theta_1)}{ q(\theta_2|\theta_1)}  \nonumber \\
&-& \sum_j \xi_j^{(2)}
\left( \int_{0}^{\infty} d \theta_2 \, G_{2j}(\theta_2)
\int d \theta_1 \, p_1(\theta_1) \, p(\theta_2|\theta_1)
 - C_{2,j} \right)
\eea
 where $ \xi_j^{(2)} $ are Lagrange multipliers enforcing the constraints
 in \eqref{constraint2}.
 Minimizing this expression with respect to $ p(\theta_2| \theta_1) $, we obtain
 the conditional transition probability
\bea
\label{secondstep}
p(\theta_2|\theta_1) &=& \frac{1}{Z_2(\theta_1, \xi^{(2)})}
q(\theta_2|\theta_1)
e^{ \sum_j \xi_j^{(2)} G_{2j}(\theta_2) }, \nonumber \\
Z_2(\theta_1, \xi^{(2)}) &=& \int_{0}^{\infty} d\theta_2 \,
q(\theta_2|\theta_1) e^{ \sum_j \xi_j^{(2)} G_{ij}(\theta_2)}.
\eea
Note that $ p(\theta_2 | \theta_1) < 0 $ if $ \theta_2 < \theta_1 $ (i.e., our ``true''
time-dilation process is a valid subordinator) as long as our prior model
$ q(\theta_2 | \theta_1) $ is a valid subordinator.

Substituting \eqref{secondstep} into \eqref{Lagrange2} (and flipping the sign to
convert a maximization problem to a minimization problem), we obtain the
following function $ U(\xi^{(2)}) $ (sometimes referred to as a  potential function):
\beq
\label{Potential}
U(\xi^{(2)}) =  \int d \theta_1 \, p(\theta_1) \,
\log Z_2(\theta_1, \xi^{(2)}) - \sum_{j}  \xi_j^{(2)} C_{2j}
\eeq
The problem of
computation of the
Lagrange multipliers $ \xi_j^{(2)} $ is now reduced to minimizing \eqref{Potential},
which again amounts to a convex optimization problem
in dimension equal to the number of option quotes for maturity $ t_2 $.


For a multi-period setting with more than two nodes on a time line, the above scheme is applied recursively. Let $ t_1,t_2,\ldots, t_N $ be nodes on the time line. We first solve the problem for the pair $ t_1,t_2 $ as described above. Using these results, we next calculate marginal probabilities $ f \left( \theta_2 \right) $ using the Chapman-Kolmogorov equations. Now the problem for the pair of times $ t_2, \, t_3 $ is treated in the exact same manner as above. We then move to the pair $ t_3,t_4 $, etc. As a result, we end up with an implied discrete-valued process for $ \theta_t $ on a discrete timeline $ t_1,t_2,\ldots, t_N $. Derivatives pricing with this framework can be done using the standard backward induction method.

\section*{Acknowledgments}
We would like to thank Gene Cohler for support and interest in this work, as well as for asking stimulating questions that have largely inspired this research. It is our pleasure to thank Peter Carr, Gene Cohler, Leonid Malyshkin, Nicolas Victoir and Olivier Vigneron for useful discussions, and Nic Trainor for editing the manuscript.


\appendixpage

\appendix
\section{2D Markov Generator on a Nonuniform Grid \label{ApNUG}}
In this appendix, we discuss how to construct the 2D Markov chain generator $A$
in \eqref{empir_matrix} using a nonuniform grid in \eqref{A_1_elements}.

To simplify notation, we introduce $h_{k,i} = \Delta z_{k,(i,i-1)}, h^+_{k,i} = \Delta z_{k,(i+1,i)}, \ k=1,2, \ i = 0,\dots,p_k$, where $p_1, p_2$ are the upper boundary of our discrete grid in the first and second dimensions. The central difference approximation of the second derivatives reads
\begin{align} \label{cd}
\left. \frac{\partial^2 V}{\partial z_1^2} \right|_{ij}  &=  \delta^-_{1,i} V_{i-1,j} + \delta^0_{1,i} V_{i,j} + \delta^+_{1,i} V_{i+1,j} +
O \left(h_{1,i}^2) \right) + O \left((h^+_{1,i})^2) \right) + O \left(h_{1,i} h^+_{1,i}) \right),\\
\left. \frac{\partial^2 V}{\partial z_2^2} \right|_{ij}  &=  \delta^-_{2,j} V_{i,j-1} + \delta^0_{2,j} V_{i,j} + \delta^+_{2,j} V_{i,j+1} +
O \left(h_{2,i}^2) \right) + O \left((h^+_{2,j})^2) \right) + O \left(h_{2,j} h^+_{2,j}) \right), \nonumber
\end{align}
\noindent where
\begin{equation*} \label{cd_coeff}
\delta^-_{k,i} = \frac{2}{h_{k,i} (h_{k,i} + h^+_{k,i})}, \quad
\delta^0_{k,i} = -\frac{2}{h_{k,i} h^+_{k,i}}, \quad
\delta^+_{k,i} = \frac{2}{h^+_{k,i} (h_{k,i} + h^+_{k,i})}.
\end{equation*}
At the boundaries these coefficients are $\delta^-_{k,i} = 0, \ i=1$ and $\delta^+_{k,i} = 0, \ i = p_k$.

For the mixed derivative we take noncentral differences to preserve nonnegativity
\begin{equation*}
\left. \frac{\partial^2 V}{\partial z_1 \partial z_2} \right|_{ij} = \sum_{m=j-1}^{j+1} \left[
\gamma^+_{i,m} V_{i+1,m} + \gamma^0_{i,m} V_{i,m} + \gamma^-_{i,m} V_{i-1,m}\right] + R_{ij}.
\end{equation*}
\noindent For $\rho \ge 0$ one has $R_{ij} =  O(h_{1,i}h^+_{2,j}) + O(h_{1,i}h^+_{1,i})$ and
\begin{align*}
\gamma^-_{i,j-1} &= \frac{1}{h_{1,i} h_{2,j}}, \quad
\gamma^-_{i,j} = - \gamma^-_{i,j-1} - \gamma^-_{i,j+1}, \quad
\gamma^-_{i,j+1} = \frac{1}{(h_{1,i}+ h^+_{1,i})h^+_{2,j}}, \nonumber \\
\gamma^0_{i,j-1} &= -\frac{1}{h_{1,i} h_{2,j}}, \quad
\gamma^0_{i,j} = -\gamma^0_{i,j-1} -\gamma^0_{i,j+1}, \quad
\gamma^0_{i,j+1} =  -\frac{1}{h^+_{1,i} h^+_{2,j}},\nonumber \\
\gamma^+_{i,j-1} &= 0, \quad
\gamma^+_{i,j} = - \gamma^+_{i,j+1}, \quad
\gamma^+_{i,j+1} = \frac{h_{1,i} }{h^+_{1,i} \left(h_{1,i}+h^+_{1,i}\right) h^+_{2,j}}. \nonumber
\end{align*}
\noindent For $\rho < 0$ we find $R_{ij} = O(h_{1,i}h^+_{2,j}) + O(h^+_{1,i} h^+_{2,j}) + O(h_{1,i}h^+_{1,i})$ and
\begin{align*}
\gamma^-_{i,j-1} &= 0, \quad
\gamma^-_{i,j} = - \gamma^-_{i,j+1}, \quad
\gamma^-_{i,j+1} = -\frac{h^+_{1,i} }{h_{1,i} \left(h_{1,i}+h^+_{1,i}\right) h^+_{2,j}}, \nonumber \\
\gamma^0_{i,j-1} &= \frac{1}{h_{2,j} h^+_{1,i}}, \quad
\gamma^0_{i,j} = -\gamma^0_{i,j-1} -\gamma^0_{i,j+1}, \quad
\gamma^0_{i,j+1} =  \frac{1}{h_{1,i} h^+_{2,j}}, \nonumber \\
\gamma^+_{i,j-1} &= -\frac{1}{h_{2,j} h^+_{1,i}}, \quad
\gamma^+_{i,j} = - \gamma^+_{i,j-1} - \gamma^+_{i,j+1}, \quad
\gamma^+_{i,j+1} = -\frac{1}{\left(h_{1,i}+h^+_{1,i}\right) h^+_{2,j}}. \nonumber
\end{align*}

Using this in \eqref{Markov_generator} and regrouping terms, we obtain
\[
\left( \mathcal{L} V(z) \right)_{ij} = \sum_{k,m = \{ -1,0,1 \} } a_{ij|i+ k,j+m} V_{i+k,j+m},
\]
where the following notation is used :
\begin{align*}
a_{ij | i+1,j} &=  \frac{1}{2} s_{ij} \sigma_1 \left( \sigma_1 \delta^+_{1,i} + \rho \sigma_2 \gamma^+_{i,j} \right),  & \qquad
a_{ij | i-1,j} &=  \frac{1}{2} s_{ij} \sigma_1 \left( \sigma_1 \delta^-_{1,i} + \rho \sigma_2 \gamma^-_{i,j} \right) \\
a_{ij | i,j+1} &=  \frac{1}{2} s_{ij} \sigma_2 \left( \sigma_2 \delta^+_{2,j} + \rho \sigma_1 \gamma^0_{i,j+1} \right),  & \qquad
a_{ij | i,j-1} &=  \frac{1}{2} s_{ij} \sigma_2 \left( \sigma_2 \delta^-_{2,j} + \rho \sigma_1 \gamma^0_{i,j-1} \right) \\
a_{ij | ij} &=  - \frac{1}{2} s_{ij} \left( \sigma_1^2 \delta^0_{1,i} + \rho \sigma_1 \sigma_2 \gamma^0_{i,j} + \sigma_2^2 \delta^0_{2,j}\right),  \\
a_{ij | i+1,j+1} &= \rho \sigma_1 \sigma_2 s_{ij} \gamma^+_{i,j+1}, & \qquad
a_{ij | i-1,j-1} &= \rho \sigma_1 \sigma_2 s_{ij} \gamma^-_{i,j-1}, \\
a_{ij | i+1,j-1} &= \rho \sigma_1 \sigma_2 s_{ij} \gamma^+_{i,j-1}, & \qquad
a_{ij | i-1,j+1} &= \rho \sigma_1 \sigma_2 s_{ij} \gamma^-_{i,j+1}. \\
\end{align*}
\noindent where $ s_{ij} = \left[ s(Z_t) \right]_{ij} $.

To construct a valid Markov generator, we have to make sure that all off-diagonal elements are
positive and all rows sum to zero. It is also necessary to obey the following property: if $f^n$ and $f^{n+1}$ are the state vectors at time moment $n$ and $n+1$, and $A$ is the transition matrix (i.e., $f^{n+1} = A f^n$), then to preserve positiveness of $f$, the matrix $A$ must be diagonally dominant. When applied to the above equations, these three conditions give rise to tricky dependencies between the grid steps $h_{1,i}, h^+_{1,i}, h_{2,i}, h^+_{2,i}$, which could be hard to reconcile with the usual approach of building a nonuniform grid based on expected values of model parameters. One possible approach that escapes the need to deal with exceedingly complicated constraints on the grid steps could be to use a nonuniform grid in one direction and a uniform grid in the other direction.\footnote{This is similar to building space grids as a part of a FD approach to solving 2D PDEs that determine the option price under some stochastic volatility models. For more details, see, e.g., \citet{Toivanen2010}.}

\section{Random Time Change of a Continuous-Time Markov Chain \label{AppA}}
Consider a homogeneous Markov
chain with a diagonalizable generator $ A $ such that
\[
A = U D U^{-1}  \; \; , \; \; D \equiv \mathrm{diag}(d_1,d_2, \ldots,d_N),
\]
where the eigenvalues $ \{ d_i \} $ are assumed to be in a descending order.
The matrix $ U $ consists of
eigenvectors stored column-wise. For a finite-time transition matrix, we then
have
\[
P(t,T) = U e^{(T-t)D} U^{-1}.
\]
Next we make the transition matrix stochastic by introducing the random time
change $ t  \rightarrow T_t $
driven by a nonnegative stochastic process (activity rate) $ Y_t $ such that
\beq
\label{tauState}
T_t = \int_{0}^{t} Y_s ds.
\eeq
By viewing $ T_t $ as a ``true'' ``business'' or ``trading'' time as opposed
to the calendar time $ t $, the transition matrix becomes stochastic as it
now depends explicitly on $ Y_t $:
\beq
\label{state}
P_X(t,T) = U e^{D \int_{t}^{T} Y_s ds} U^{-1}.
\eeq
Consider now a Markov chain obtained by conditioning on a path of $ Y_t $.
By taking the derivative of \eqref{state} with respect to $ t $ and comparing with the
Kolmogorov equation
\begin{equation*}
\frac{ \partial P_X(t,T)}{ \partial t} = - A_X(t) P_X(t,T),
\end{equation*}
we see that the conditional on the realization of the path of $ Y_t $, our
process is given by an inhomogeneous Markov chain with  generator
\[
A_X(t) = Y_t U D U^{-1} = Y_t A.
\]

\end{document}